\def\regenfigs{0}
\documentclass[sigconf,9pt]{acmart}
\AtBeginDocument{%
  }

\setcopyright{cc}
\setcctype{by}
\acmConference[SC Workshops '25]{Workshops of the International Conference
for High Performance Computing, Networking, Storage and Analysis}{November
16--21, 2025}{St Louis, MO, USA}
\acmBooktitle{Workshops of the International Conference for High Performance
Computing, Networking, Storage and Analysis (SC Workshops '25), November
16--21, 2025, St Louis, MO, USA}
\copyrightyear{2025}
\acmYear{2025}
\acmDOI{10.1145/3731599.3767498}
\acmISBN{979-8-4007-1871-7/2025/11}

\usepackage{microtype}
\if\regenfigs1
\usepackage{tikz,pgfplots}
\usetikzlibrary{arrows.meta}
\usepgfplotslibrary{groupplots}
\usepgfplotslibrary{external}
\usepgfplotslibrary{colorbrewer}
\pgfplotsset{cycle list/Set1}
\usepgfplotslibrary{fillbetween}
\pgfplotsset{compat=1.18}
\pgfplotsset{
    tick align=outside,
    tick pos=left,
    xmajorgrids,
    x grid style={white},
    ymajorgrids,
    y grid style={white},
    axis line style={white},
    axis background/.style={fill=white!89.803921568627459!black},
    legend style={draw=none, fill=none},
    legend cell align=left,
}
\pgfkeys{/pgf/number format/.cd, 1000 sep={\,}}

\pgfplotsset{
    log x ticks with fixed point/.style={
        xticklabel={
            \pgfkeys{/pgf/fpu=true}
            \pgfmathparse{2^\tick}%
            \pgfmathprintnumber[fixed relative, precision=4]{\pgfmathresult}
            \pgfkeys{/pgf/fpu=false}
        }
    },
    log10 x ticks with fixed point/.style={
        xticklabel={
            \pgfkeys{/pgf/fpu=true}
            \pgfmathparse{10^\tick}%
            \pgfmathprintnumber[fixed relative, precision=3]{\pgfmathresult}
            \pgfkeys{/pgf/fpu=false}
        }
    },
    log y ticks with fixed point/.style={
        yticklabel={
            \pgfkeys{/pgf/fpu=true}
            \pgfmathparse{2^\tick}%
            \pgfmathprintnumber[fixed relative, precision=4]{\pgfmathresult}
            \pgfkeys{/pgf/fpu=false}
        }
    }
}
\fi

\begin{document}

\title{LAMMPS-KOKKOS: Performance Portable Molecular Dynamics Across Exascale Architectures}

\author{Anders Johansson}
\affiliation{%
  \institution{Sandia National Laboratories}
  \city{Albuquerque}
  \state{New Mexico}
  \country{USA}}
\email{ajohans@sandia.gov}

\author{Evan Weinberg}
\affiliation{%
  \institution{NVIDIA Corporation}
  \city{Santa Clara}
  \state{California}
  \country{USA}}
\email{eweinberg@nvidia.com}

\author{Christian R. Trott}
\affiliation{%
  \institution{Sandia National Laboratories}
  \city{Albuquerque}
  \state{New Mexico}
  \country{USA}}
\email{crtrott@sandia.gov}

\author{Megan J. McCarthy}
\affiliation{%
  \institution{Sandia National Laboratories}
  \city{Albuquerque}
  \state{New Mexico}
  \country{USA}}
\email{megmcca@sandia.gov}

\author{Stan G. Moore}
\affiliation{%
  \institution{Sandia National Laboratories}
  \city{Albuquerque}
  \state{New Mexico}
  \country{USA}}
\email{stamoor@sandia.gov}

\renewcommand{\shortauthors}{Johansson et al.}

\begin{abstract}
Since its inception in 1995, LAMMPS has grown to be a world-class molecular dynamics code, with thousands of users, over one million lines of code, and multi-scale simulation capabilities.
We discuss how LAMMPS has adapted to the modern heterogeneous computing landscape by integrating the Kokkos performance portability library into the existing C++ code.
We investigate performance portability of simple pairwise, many-body reactive, and machine-learned force-field interatomic potentials.
We present results on GPUs across different vendors and generations, and analyze performance trends, probing FLOPS throughput, memory bandwidths, cache capabilities, and thread-atomic operation performance. 
Finally, we demonstrate strong scaling on three exascale machines -- OLCF Frontier, ALCF Aurora, and NNSA El Capitan -- as well as on the CSCS Alps supercomputer, for the three potentials.

\end{abstract}




\maketitle

\section{Introduction} \label{sec:introduction}
Molecular dynamics is a method for studying materials and molecules by directly integrating Newton's second law to obtain trajectories that describe chemical reactions, heat transport, diffusion, and other physical phenomena that occur at the time and length scales that are within computational reach of atomistic methods.
Since the initial efforts by Rahman in 1964 \cite{rahman1964correlations},  the increase in applicability of molecular dynamics has been driven by the exponential growth in computational power that enables simulations of larger systems over longer timescales with higher accuracy \cite{Plimpton2012,JCP}.

Further progress made in the 1990s and the spread of distributed computing presented a major leap. 
The natural spatial decomposition of molecular dynamics simulations made them a prime example for adopting the new massive parallelism paradigm, as demonstrated by the first releases of still-popular software such as LAMMPS in 1995~\cite{lammps95}, as well as GROMACS~\cite{abraham2015gromacs} and NAMD~\cite{namd}. 
Over the next decade and a half, LAMMPS grew in scope and capability, from 23k lines of Fortran code in 1999, to 148k lines of C++ in 2009, but the MPI-distributed domain decomposition computing paradigm remained the same.
Then, GPUs became more general purpose with the introduction of programming models such as CUDA. Their high FLOP rate and improved power efficiency gave significant advantages over CPUs, leading to widespread adoption of heterogeneous computing clusters.

With parallelism being a major strong point of GPUs, molecular dynamics codes with their large number of force calculations once again became an early adopter of the new heterogeneous computing paradigm.
New codes were developed such as HOOMD-blue~\cite{anderson2020hoomd} and HAL's MD~\cite{halmd}, while existing codes raced to adopt the new technology as quickly and efficiently as possible. For LAMMPS, whose modular \emph{package} system (described in section~\ref{sec:packages} below) allows development of different approaches simultaneously, two competing strategies soon emerged.

The GPU \emph{package} was released as part of LAMMPS in 2010 and took the common approach of simply offloading the force calculation, which is typically the most computationally expensive step. Nearly all other kernels run on the host CPU. This requires frequent data copies between host and device in every timestep.
While reasonable speedups were achieved, particularly when launching multiple MPI tasks per GPU, this method has clear drawbacks given the limited transfer speed and high latency between the separate memories of the CPU and the GPU.

This paper instead focuses on the second strategy: the KOKKOS \emph{package} in LAMMPS, which uses the Kokkos \emph{library} \cite{trott2021kk3} to achieve performance portability.
The KOKKOS package (released in 2014) has design origins in the USER-CUDA package developed by Trott~\cite{phd-trott,lammpscuda}, which was released as part of LAMMPS in 2011.
The USER-CUDA package attempted to be ``GPU resident'', meaning as many kernels as possible run on the GPU. This avoids copying data between GPU and CPU at every timestep, leading to performance benefits. However, data transfer is still necessary to maintain compatibility when using functionality not yet ported to GPUs. The KOKKOS package brought the additional promise of keeping LAMMPS vendor-agnostic, which has proven vital given the modern introduction of large-scale computing resources using GPUs from AMD, Intel, and NVIDIA. 

In this paper, we discuss the current structure of LAMMPS and how the optional KOKKOS package provides seamless performance portability for the user when enabled.
We include a comprehensive set of benchmarks for comparing how NVIDIA GPU performance has improved over hardware generations (V100-16GB-SXM3, A100-40GB-SXM4, and H100-HBM3-SXM5), and how LAMMPS benchmarks perform on three existing exascale systems: Frontier (AMD MI250X)~\cite{frontier}, El Capitan (AMD MI300A)~\cite{elcapitan} and Aurora (Intel Data Center GPU Max 1550 -- subsequently referred to as PVC)~\cite{aurora}, along with the Alps supercomputer (NVIDIA Grace-Hopper Superchip -- subsequently referred to as GH200)~\cite{alps} and NVIDIA's Eos (NVIDIA DGX H100 Superpod)~\cite{eos}. For reference, we include a comparison of different hardware in table~\ref{tab:gpu-specs}.

Both AMD MI250X and Intel PVC contain two logical GPUs, and in this work single GPU results use a single ``GCD'' (GPU compute die) for MI250X and a single ``stack'' for PVC. We note that for single GPU results we focused on NVIDIA H100, while multi-GPU results used NVIDIA GH200. In the context of this work the differences are minimal; we quantify this in section~\ref{sec:eos-alps}.

We present case studies on the Lennard-Jones, SNAP, and ReaxFF potentials, and a discussion of how LAMMPS keeps up with current developments in the field of molecular dynamics such as machine-learning-based potentials.

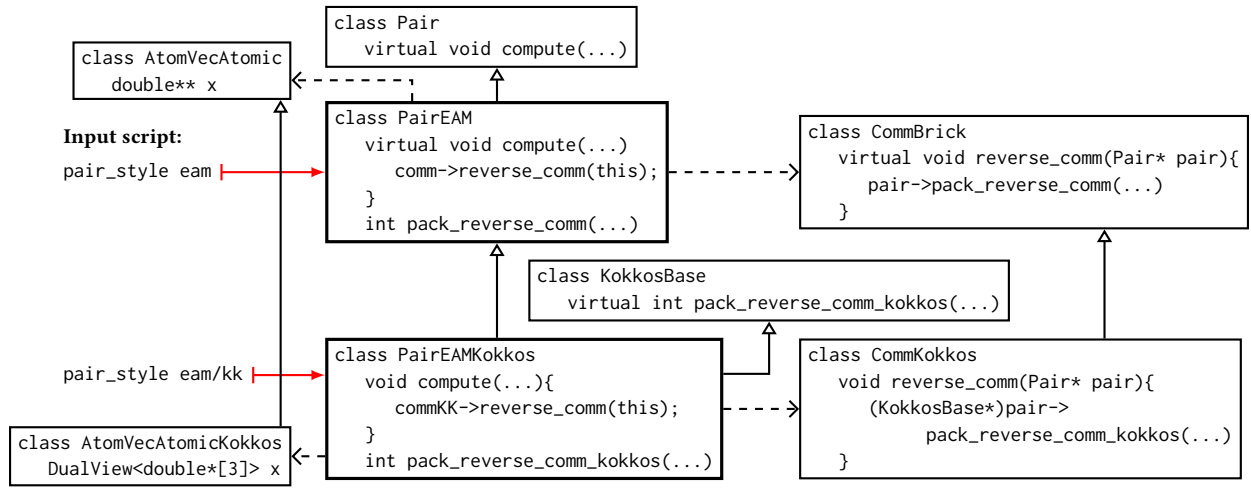
\begin{figure*}[tbp]
\centering
\if\regenfigs1
\begin{tikzpicture}[x=0.9cm,y=0.9cm,thick,>=latex,font=\small]
    \draw (0,0) node[anchor=west] (paiream) {\texttt{pair\_style eam}};
    \draw (0,-3) node[anchor=west] (paireamkk) {\texttt{pair\_style eam/kk}};
    \draw (0,0.5) node[anchor=west] (input) {\textbf{Input script:}};

    \draw (4,0) node[very thick,draw,anchor=west,align=left] (pair) {\texttt{class PairEAM}\\ \hspace{0.4cm}\texttt{virtual void compute(...)}\\
        \hspace{0.8cm}\texttt{comm->reverse\_comm(this);}\\
        \hspace{0.4cm}\texttt{\}}\\
    \hspace{0.4cm}\texttt{int pack\_reverse\_comm(...)}};
    \draw (4,-3.5) node[very thick, draw,anchor=west,align=left] (pairkk) {\texttt{class PairEAMKokkos}\\
        \hspace{0.4cm}\texttt{void compute(...)\{}\\
        \hspace{0.8cm}\texttt{commKK->reverse\_comm(this);}\\
        \hspace{0.4cm}\texttt{\}}\\
        \hspace{0.4cm}\texttt{int pack\_reverse\_comm\_kokkos(...)}};

    \draw[<-|, red] (pair.west |- paiream.east) -- (paiream.east); 
    \draw[<-|, red] (pairkk.west |- paireamkk.east) -- (paireamkk.east); 

    \draw (4,2.0) node[draw,anchor=west,align=left] (pairbase) {\texttt{class Pair}\\ \hspace{0.4cm}\texttt{virtual void compute(...)}};
    \draw (7,-1.75) node[draw,anchor=west,align=left] (kkbase) {\texttt{class KokkosBase}\\ \hspace{0.4cm}\texttt{virtual int pack\_reverse\_comm\_kokkos(...)}};

    \draw[{Triangle[open]}-] (pairbase.south -| pair.north) -- (pair.north);
    \draw[-{Triangle[open]}] (pairkk.north -| pair.south) -- (pair.south);
    \draw[{Triangle[open]}-]  (kkbase.south) |- (pairkk.10);

    \draw (3.5,1.5) node[draw, align=left, anchor=east] (atom) {\texttt{class AtomVecAtomic}\\ \hspace{0.4cm}\texttt{double** x}};  
    \draw (3.5,-4.2) node[draw, align=left, anchor=east] (atomkk) {\texttt{class AtomVecAtomicKokkos}\\ \hspace{0.4cm}\texttt{DualView<double*[3]> x}};  
    
    \draw[{Triangle[open]}-] (atom.345 -| atomkk.13) -- (atomkk.13);
    \draw[dashed,-{Straight Barb[]}] (pair.140) |- (atom.355);
    \draw[dashed,-{Straight Barb[]}] (pairkk.west |- atomkk.east) -- (atomkk.east);
    
    \draw (11,0) node[draw, align=left, anchor=west] (comm) {\texttt{class CommBrick}\\
        \hspace{0.4cm}\texttt{virtual void reverse\_comm(Pair* pair)\{}\\
        \hspace{0.8cm}\texttt{pair->pack\_reverse\_comm(...)} \\
        \hspace{0.4cm}\texttt{\}}
    };
    \draw (11,-3.5) node[draw, align=left, anchor=west] (commkk) {\texttt{class CommKokkos}\\
        \hspace{0.4cm}\texttt{void reverse\_comm(Pair* pair)\{}\\
        \hspace{0.8cm}\texttt{(KokkosBase*)pair->} \\ 
        \hspace{1.5cm} \texttt{pack\_reverse\_comm\_kokkos(...)} \\
        \hspace{0.4cm}\texttt{\}}
    };  
    \draw[dashed,-{Straight Barb[]}] (pair.east |- comm.west) -- (comm.west);
    \draw[dashed,-{Straight Barb[]}] (pairkk.east |- commkk.west) -- (commkk.west);
    \draw[{Triangle[open]}-] (comm.320 -| commkk.40) -- (commkk.40);
\end{tikzpicture}
\else
\includegraphics{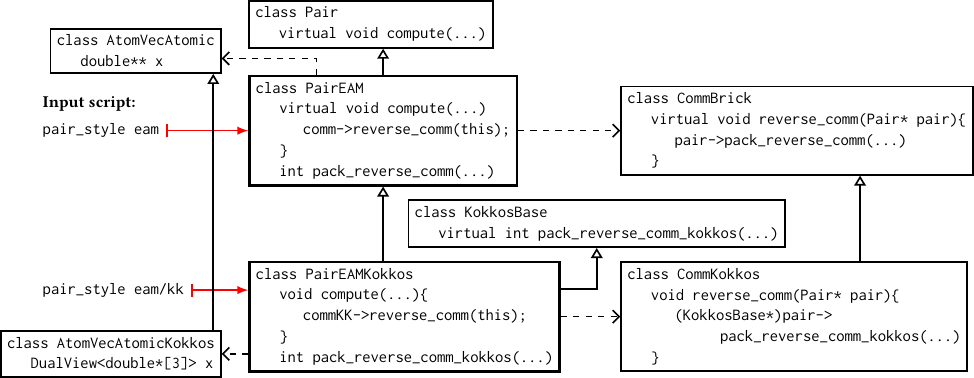}
\fi
\caption{Mapping (red arrows) from user input script commands to LAMMPS class hierarchy and how the classes interact within and outside of the KOKKOS package. The pair styles operate on atomic data stored as pointers in \texttt{AtomVecAtomic} and \texttt{Kokkos::DualView}s in \texttt{AtomVecAtomicKokkos}, with the pointer in the former aliased to the CPU mirror of the latter. The EAM pair style requires additional communication, which is performed with calls to the LAMMPS communication classes. Following Unified Modeling Language (UML) conventions, solid black arrows indicate inheritance, while dashed arrows indicate usage.}\label{fig:flow}
\end{figure*}

\begin{table}[tbp] 
  \centering        
  \caption{GPU architecture properties, listing HBM memory bandwidth (``BW'') and capacity, floating point throughput (FP64, excludes matrix multiplication hardware), and cache size (single numbers are used for combined caches, two numbers for separate hardware and software managed caches). For AMD MI250X and Intel PVC a single logical GPU is listed, not the full package.}
  \label{tab:gpu-specs} 
  \begin{tabular}{|l|l|l|l| l |} 
    \hline 
    GPU & BW & Capacity & FP64 & L1 + Shared \\ \hline 
    NVIDIA V100 & 0.9 TB/s & 16 GB & 7.8 TF & 128 kB \\
    NVIDIA A100 & 1.5 TB/s & 40 GB & 9.7 TF & 192 kB\\
    NVIDIA H100 & 3.3 TB/s & 80 GB & 34 TF & 256 kB\\
    NVIDIA GH200 & 4.0 TB/s & 96 GB & 34 TF & 256 kB\\
    AMD MI250X/2 & 1.6 TB/s & 64 GB & 24 TF & 16 + 64 kB\\
    AMD MI300A & 5.3 TB/s & 128 GB & 61 TF & 32 + 64 kB\\
    Intel PVC stack & 1.6 TB/s & 64 GB & 26 TF & n/a + 128 kB\\
    \hline 
  \end{tabular}
\end{table}


\section{LAMMPS basic code structure}\label{sec:lammps-structure}
LAMMPS is one of the leading molecular dynamics simulation codes, in part because of the flexibility and collection of capabilities that have grown through over 30 years of development and user contributions \cite{kohlmeyer2025usrse,lammps22}. 

When attempting to make LAMMPS performance portable through use of the Kokkos library, it was important to retain this flexibility and ensure it remained straightforward for users to develop their own LAMMPS extensions without knowledge of Kokkos abstractions.
In this section, we describe the basic code structure of LAMMPS for providing and exposing the different functionality, and how the Kokkos functionality builds upon and remains compatible with this structure.

\subsection{From user input to C++} \label{sec:input}
Users interact with LAMMPS through input scripts, which create the initial structure of atoms, assign an \textit{interatomic potential} that will provide the atomic forces from atomic positions, set up simulation modifiers and output diagnostics, and finally start the simulation.
Each step is executed using one or more of a varied set of LAMMPS \emph{commands}.
There are two main types of commands in the scripting language; those that execute immediately, such as \texttt{read\_data} for reading an atomic structure from a file, and those that are executed repeatedly during the subsequent simulation, such as \texttt{fix rigid}, which will force a specified subset of the atoms to behave as a rigid body.

The LAMMPS input script parser contains a map from each command to the corresponding C++ class, where the map is created by a macro called in the header file containing the declaration of the class.
When the command is parsed, an instance of the C++ class is created.
For commands that execute immediately, LAMMPS calls the constructor and the \texttt{command} method of the instance, before discarding the instance and moving on to the next command.
For all other commands, the instance is stored in a list of objects whose methods are called at specified intervals during subsequent simulations.

\subsection{Styles}\label{sec:styles}
These persistent classes are referred to as \emph{styles}, which belong to different categories described more closely in the LAMMPS reference paper~\cite{lammps22}.
LAMMPS contains hundreds of different styles, in different categories. ``Fix'' styles have methods that are called at arbitrary points and intervals during the simulation to either modify the trajectory of the simulation or generate output, while ``compute'' styles can be viewed as restricted versions of fixes that only produce output accessible to the user within the input script, without modifying the state of the system.
The case studies presented in section~\ref{sec:casestudies} are examples of \emph{pair styles}, with each pair style implementing a different model for the potential energy, often referred to as simply ``potentials''.
At each timestep, the derivatives of the selected potential with respect to atomic coordinates provide the forces that are used for time integration, making pair styles the most important category of styles since they are typically the most expensive part of a simulation.
The examples in section~\ref{sec:casestudies} span a range of complexities, from a simple pairwise potential to a state-of-the-art machine-learning-based potential.

\section{LAMMPS-KOKKOS package}\label{sec:lmpkk}

The KOKKOS package in LAMMPS was one of the first serious applications for Kokkos, with some of Kokkos's design features directly inspired by use-cases from LAMMPS -- including in the area of data structures and parallel execution.
In this section, we describe how Kokkos is integrated to seamlessly enable performance portability without losing access to preexisting non-Kokkos functionality.

\subsection{Packages and suffixes}\label{sec:packages}
By default, LAMMPS is compiled with a restricted set of functionality and a limited number of styles, without the optional KOKKOS package (i.e., only pure C++ code using MPI), in order to simplify and accelerate compilation. 
For users that require additional capabilities, the code is organized in \emph{packages}, which provide either \emph{additional} or \emph{replacement} functionality.
Examples of packages that provide \emph{additional} functionality include \texttt{MANYBODY}, for many-body potentials such as the Embedded Atom Method (EAM) \cite{EAM}; \texttt{MOLECULE}, for bonded interactions; and \texttt{KSPACE}, for long-range interactions that require Fourier transforms and calculations in reciprocal space (``k-space'').
The ability to add and remove packages at will at compile-time gives LAMMPS users excellent flexibility while also ensuring code modularity and ease of maintenance.

The ``accelerator'' packages, on the other hand, provide \emph{replacement} functionality in the form of accelerated styles, i.e., classes that provide equivalent functionality but improve computational performance on different hardware.
This paper describes the \texttt{KOKKOS} package, which provides alternate versions of many classes that are accelerated by the Kokkos library for performance portability.
As shown in figure~\ref{fig:flow}, the \text{KOKKOS} package contains a \texttt{PairEAMKokkos} class for the EAM potential that is equivalent to the base \texttt{PairEAM} class from which it inherits, but its computational routines implement Kokkos library abstractions.

At runtime, the accelerated versions need to be selected through the map from input script command to C++ class described in section~\ref{sec:input} and figure~\ref{fig:flow}.
Each accelerated style is registered and added to the map using the same macro as the non-accelerated style, with the convention that the accelerated styles append a package-specific \emph{suffix}, namely \texttt{/kk} for the \texttt{KOKKOS} package.
The user can then selectively choose the accelerated version of a style using, for example, the \texttt{eam/kk} pair style instead of \texttt{eam} in their input script, or they can use the Kokkos-accelerated version of all possible styles used in the input script by specifying a global suffix.
Through careful design of the data structures and data movement patterns described in this section, the user is allowed to accelerate the parts of their simulation for which accelerated styles are available, without losing access to those for which no Kokkos support currently exists.

\subsection{Data Structures}
The primary data structure in Kokkos is the \texttt{Kokkos::View}, a structure that can represent multi-dimensional arrays, and encodes both data layout and data accessible through type mechanism via its \texttt{Layout} and \texttt{MemorySpace} template parameters.
\texttt{Kokkos::View} strongly informed the design of the ISO C++23 \texttt{std::mdspan} capability.
The Kokkos variants of styles in LAMMPS generally contain host and device variants of data encapsulated in a \texttt{Kokkos::DualView}. 
This data structure contains a compatible \texttt{Kokkos::View} for both a host (i.e.~CPU) and a device (i.e.~GPU) memory space to help with unstructured data synchronization.
In particular, it has functionality to keep track of when data was modified, and thus when data has to be synced.
The classic non-Kokkos data fields in LAMMPS are initialized to alias the allocations underlying the host \texttt{View} of the Kokkos data structures via Kokkos's \texttt{View} interoperability with raw pointers.

Since the KOKKOS package is designed such that both Kokkos and non-Kokkos styles can be used together within a single LAMMPS input script, managing data transfers poses a particular challenge.
To address this, every style (fix, pair style, etc.) has a set of flags that indicate whether the style will read and/or modify a specific data field.
These flags are then used to determine whether to call the synchronization functionality of \texttt{DualView}, which keeps track of when it was last synced.
Consequently, simply calling sync inside a LAMMPS style when it needs to access a data field will only incur the overhead of actual memory transfer between host and device if the data was last modified in the other (non-accessible) memory space.
Thus, no global knowledge of the required data transfer patterns is necessary.

If LAMMPS is configured for a pure host build, \texttt{DualView}'s synchronization mechanisms effectively become inactive, and thus the built-in data transfer functionality does not incur any overhead.

Another higher level data structure relied upon by LAMMPS is \texttt{ScatterView}.
This data structure was designed to handle unstructured accumulation of data from multiple threads in a way that write conflicts are avoided.
It can transparently swap between using atomic operations, a data duplication strategy, or even simple sequential accumulation in case of non-threaded execution.
This is important to handle the different amounts of concurrency and atomic operation throughput on various architectures.
On CPUs, data duplication with a subsequent combining step is often the most effective way to deal with write conflicts, while on GPUs data duplication is infeasible due to the large number of active threads (O(100,000)) and thus atomic operations need to be used. 

\subsection{Execution Control}\label{sec:exe-control}

Nearly every Kokkos-based style in LAMMPS is templated on the device type and is instantiated for both the default host and device execution space simultaneously.
This dual instantiation pattern enables users to request a Kokkos-based style for either host CPU or GPU device at runtime from the input script through the \texttt{/kk/host} suffix and the \texttt{/kk/device} suffix (equivalent to \texttt{/kk}, see section~\ref{sec:packages}), which can greatly improve execution control.
One example of potential benefits is in the communication phase, where, depending on the size of the problem, the availability of GPU-aware MPI, and the actual hardware, it may be more performant to keep all communication routines (packing, unpacking, sending data) on host, or execute it on the device.

Kokkos parallel execution is controlled via execution patterns and execution policies.
The primary patterns are \texttt{parallel\_for}, \texttt{parallel\_reduce} and \texttt{parallel\_scan} which execute a callable for each index in an iteration space, and if applicable, combine results via reductions.
The main policies are \texttt{RangePolicy}, \texttt{MDRangePolicy} and \texttt{TeamPolicy}.
\texttt{RangePolicy} and \texttt{MDRangePolicy} simply iterate over a 1D or multi-dimensional range of indices, respectively.
The latter also enables tiling, which can be beneficial to achieve better cache locality in multi-dimensional loop patterns.

\texttt{TeamPolicy} enables the use of hierarchical (nested) parallelism, i.e.~exposing concurrency in non-tightly nested loops.
Specifically it exposes a concept of thread teams, which can collaboratively execute nested parallel loops, as well as vector-level parallelism as a third nesting level using \texttt{TeamThreadRange}, \texttt{TeamVectorRange}, and \texttt{ThreadVectorRange} respectively.
It also enables the use of scratch memory, which on GPUs can be mapped to software-managed caches such as NVIDIA's ``shared memory'', a topic which will be explored in more depth in section~\ref{sec:casestudies-cache}.

Execution policies are templated on the execution (or device) space, enabling fine-grained control of where parallel sections of code are dispatched as well.
Using these strongly-typed execution spaces, one can achieve algorithmic specialization or runtime parameter tuning for specific architectures via template specialization, C++ \texttt{if constexpr} conditions, or function overload resolution.
This can be necessary due to the vast difference in available concurrency and other hardware characteristics to achieve optimal performance.
However, our general experience is that full algorithmic specialization for architectures is rarely needed (perhaps for less than 5\% of the algorithms).
Even in these cases, Kokkos allows one to write both algorithms using the same programming paradigm and the same APIs.
For more details please refer to \cite{trott2021kk3} and \cite{edwards2014}.

\section{Case studies}\label{sec:casestudies}
In this paper, we focus on three case studies to demonstrate strategies for accelerating interatomic potentials in LAMMPS.

The first case study is the \textbf{Lennard-Jones (LJ) potential} \cite{LJ}, which is a simple pairwise potential that models the force between two atoms as a function of their distance. 

For a general pairwise potential, the total energy of the system is a sum over all pairs of atoms that are within a specified cutoff distance. 
The interatomic forces are then the pairwise sum of the derivative of the potential energy with respect
to position. 
The summation over pairs of close atoms is enabled by the neighbor list constructed and periodically updated by LAMMPS according to the algorithms described in the LAMMPS reference paper~\cite{lammps22}.

The LJ potential energy is given as
\begin{equation}
    E = \sum_{i<k, r_{ik}<r_\text{c}} 4\varepsilon \left[\left(\frac{\sigma}{r_{ik}}\right)^{12} -\left(\frac{\sigma}{r_{ik}}\right)^6\right] = \sum_{i<k, r_{ik}<r_\text{c}} U^{(2)}~(r_{ik}),\label{eq:pairwise}
\end{equation}
with \(r_{ik}\) the distance between atoms \(i\) and \(k\), $r_\text{c}$ is the interaction cutoff distance that may vary between atomic systems, and $\varepsilon$ and $\sigma$ are coefficients unique to LJ that may also vary between given atomic systems. 
LJ refers to a specific functional form of the more general idea of a two body potential $U^{(2)}(r_{ik})$. Different systems call for different functional forms. 
LJ is good for modeling noble gases such as argon; electrically charged systems may add the Coulomb potential as well.


While pairwise potentials are conceptually simple, they are an important starting point for modeling atomic interactions.
They are also often a constituent of more complex, multi-body potentials that are used to model more complicated systems that cannot be described by simple pairwise interactions alone.
In a similar vein, the simplicity of pairwise potentials makes them an important starting place for porting efforts before moving on to conceptually and algorithmically complex potentials.
We investigate the implementation of pairwise potentials in LAMMPS
in greater detail in section~\ref{sec:casestudies-pairwise}.


The second case study is the \textbf{reactive potential ReaxFF}~\cite{van2001reaxff}, which was designed for studying chemical processes such as combustion and catalytic reactions. ReaxFF is an \textit{empirical} model explicitly designed to support physical phenomena such as bond formation and breaking, polarization, and charge transfer effects. This explicit empirical design is in contrast to machine-learned potentials which, when properly designed, could ``learn'' these dynamics.

The energy of the ReaxFF potential is a sum of traditional pairwise non-bonded interactions in which all neighboring atoms interact, empirical descriptions of two-, three-, and four-body bonded interactions in which only bonded sets of atoms interact, and terms encoding polarization and charge transfer effects. 
The charge transfer effects are encoded in a \textit{separate} process known as the charge equilibration (QEq) method, defined by minimizing the total electrostatic energy of the system while maintaining charge conservation. 
Each portion of the energy is based on empirical observations of real dynamical systems.
We investigate implementing and optimizing the critical portions of the ReaxFF potential in the KOKKOS package in section~\ref{sec:casestudies-reaxff}.

The third case study is the \textbf{machine-learning-based potential SNAP (Spectral Neighbor Analysis Potential)}~\cite{thompson2015spectral}.
Machine-learned potentials generally maintain the assumption of local interactions, but allow a more complex \textit{and} general functional form that can consume information about the entire atomic neighborhood of an atom to produce the per-atom energy,
\( 
    E_i = E_{\text{ML}}\left(\left\{\vec{r}_{ik} \| r_{ik} < r_{\text{c}}\right\}\right),
\) 
which is then summed to the total energy.

One choice of a general functional form is a truncated spectral decomposition. 
For SNAP, the relative distances between atoms $\vec{r}_{ik}$ are mapped onto a hypersphere, and then the total neighborhood is decomposed in a hyperspherical harmonic basis (Wigner U-matrices $\bf{u}_j$). 
This is analogous to a spherical harmonic basis ($Y_{lm}$) on spheres or a Fourier (k-space) basis (sine and cosine functions) on the real line. 

SNAP is defined by a linear combination of triple products and is a ``machine learning'' potential because it ``learns'' the coefficients of this linear combination. 
The relative advantage of SNAP (and other machine-learned potentials) to empirical, fixed functional forms such as Lennard-Jones and ReaxFF is both its generality and its ability to train the weights of this general form. 
As an example, linear SNAP can encode general four-body interactions. 
This means that, when properly trained, it can \textit{learn} the type of four-body physics encoded in ReaxFF. 
In contrast, ReaxFF, having a fixed, empirical form, cannot be fit to a more general type of four-body interaction. 

We defer to \cite{thompson2015spectral} for the more granular details of SNAP. 
We will introduce the minimum required mathematical formalism and then discuss the challenges and optimizations of the Kokkos implementation of SNAP in section~\ref{sec:casestudies-snap}.


\subsection{Simple Pairwise Interatomic Potentials (Lennard-Jones)}\label{sec:casestudies-pairwise}

\begin{figure}[tbph]
\if\regenfigs1
\begin{tikzpicture}[overlay=false,>=latex]
    \begin{groupplot}[
            /tikz/overlay=false,
            /tikz/thick,
            group style={
                group size=1 by 2,
                vertical sep=0.7cm,
                xlabels at=edge bottom,
                ylabels at=edge left,
            },
            width=\linewidth,
            ylabel near ticks,
            xlabel={\# atoms},
            xmode=log,
            log basis x=10,
            clip=false,
            title style={yshift=-0.5cm, xshift=-2.8cm,anchor=west},
            ylabel style={xshift=1.6cm},
            height=2.1in,
        ]
            \nextgroupplot[
                legend style={at={(1.0,0.3)},anchor=east},
                legend columns=3,
                title={a) Neighbor threading},
                transpose legend,
                ymode=log,
            ]
            \addlegendimage{empty legend};
            \addlegendentry{\hspace{-0.55cm}H100};
            \pgfplotsset{cycle list/Reds-5}
            \pgfplotsset{cycle list shift=3}
            \addplot+[mark=*]  table {results/pair/thread_h100_off.txt};
            \addlegendentry{off};
            \addplot+[mark=*]  table {results/pair/thread_h100_on.txt};
            \addlegendentry{on};
            \addlegendimage{empty legend};
            \addlegendentry{\hspace{-0.65cm}MI250X};
            \pgfplotsset{cycle list/Blues-5}
            \pgfplotsset{cycle list shift=0}
            \addplot+[mark=square*]  table {results/pair/thread_mi250x_off.txt};
            \addlegendentry{off};
            \addplot+[mark=square*]  table {results/pair/thread_mi250x_on.txt};
            \addlegendentry{on};
            
            \nextgroupplot[
                legend style={at={(0.0,0.6)},anchor=west},
                legend columns=4,
                title={b) Neighbor list style and Newton},
                transpose legend,
                ylabel={Speed [M\,atom-steps/s/node]},
            ]
            \addlegendimage{empty legend};
            \addlegendentry{\hspace{-0.55cm}H100};
            \pgfplotsset{cycle list/Reds-5}
            \pgfplotsset{cycle list shift=2}
            \addplot+[mark=*]  table {results/pair/neigh_h100_full.txt};
            \addlegendentry{full};
            \addplot+[mark=*]  table {results/pair/neigh_h100_half_off.txt};
            \addlegendentry{half, off};
            \addplot+[mark=*]  table {results/pair/neigh_h100_half_on.txt};
            \addlegendentry{half, on};
            
            \addlegendimage{empty legend};
            \addlegendentry{\hspace{-0.55cm}MI250X};
            \pgfplotsset{cycle list/Blues-5}
            \pgfplotsset{cycle list shift=-1}
            \addplot+[mark=square*]  table {results/pair/neigh_mi250x_full.txt};
            \addlegendentry{full};
            \addplot+[mark=square*]  table {results/pair/neigh_mi250x_half_off.txt};
            \addlegendentry{half, off};
            \addplot+[mark=square*]  table {results/pair/neigh_mi250x_half_on.txt};
            \addlegendentry{half, on};
    \end{groupplot}
\end{tikzpicture}
\else
\includegraphics{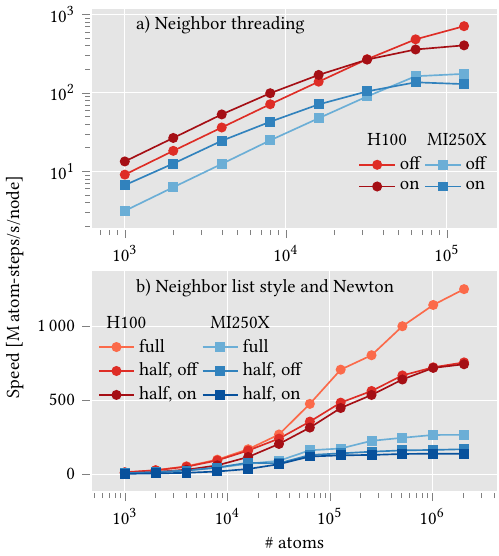}
\fi
\caption{Performance effect of different options for a simple pairwise LJ potential. a) Effect of exposing parallelism over neighbors as a function of the number of atoms. For small systems, the benefit of additional parallelism outweighs the reduced efficiency of the more complex iteration pattern. b) Effect of redundant computations vs. thread-level atomic operations. With a full neighbor list, Newton's third law is ignored, and all pair interactions are computed twice. In return, this avoids the atomic operations and, with \texttt{newton on}, additional communication required with half neighbor lists. For simple pairwise potentials, whose computational cost is low, the full neighbor list is faster.}\label{fig:pair}
\end{figure}
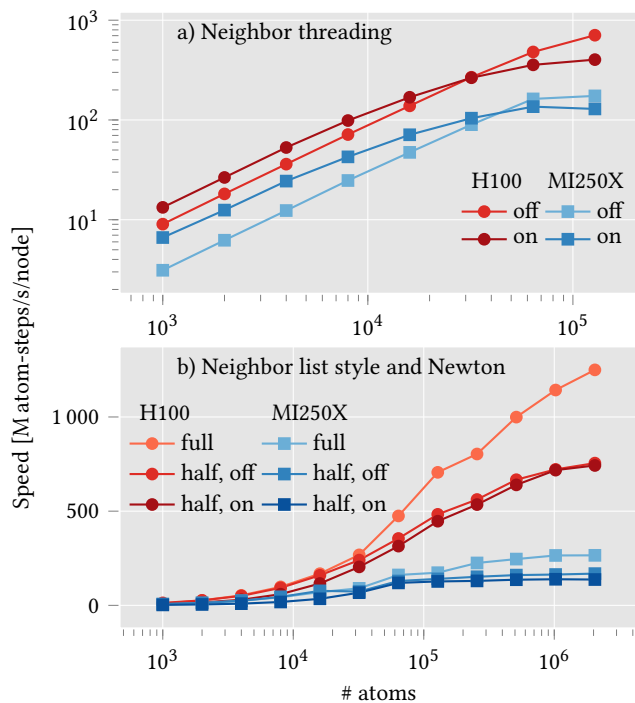

In the KOKKOS package, most two-body forces are implemented through a \texttt{pair\_kokkos} abstraction. 
Each two-body \textit{pair style} derives from a base ``\texttt{PairKokkos}'' class that contains a method defining a generic two-body potential. 
The derived class implements its own kernels that only compute the pairwise force and, if required, energy for the specific potential form. 
The base class handles all other details: neighbor list style, managing \texttt{ScatterView} objects, radial cutoff calculations, accumulating forces and energies, etc.

There are two neighbor list styles: ``half'' and ``full''.
Using ``half'' neighbor lists exploits Newton's third law, avoiding the duplicate computation of symmetric forces between two atoms.
With ``full'' neighbor lists the force of an atom \texttt{i} onto an atom \texttt{k} is computed separately from the force of the atom \texttt{k} onto the atom \texttt{i}.
For simple pairwise potentials, using a ``half'' neighbor list can cause a data write conflict between two threads when writing to the atom force array, which needs to be handled via a deconflicting approach, namely thread atomic operations or data duplication.
Data deconflicting is handled transparently by Kokkos \texttt{ScatterView} objects.
Alternatively, one can use a ``full'' neighbor list, which does not require atomics but duplicates work (every atom's force is computed twice).
Separately, LAMMPS has the option to use Newton's third law for ghost atoms (copies of atoms owned by other MPI ranks or on the other side of a periodic boundary), which reduces computation but increases the amount of communication required.
For simple pairwise potentials, typically a full neighbor list and \texttt{newton off} is better for GPUs, while a half list and \texttt{newton on} is better for CPUs (in particular in the common case when using one MPI rank per core, and not multi-threading).

Which neighbor list style to use does not have a one-size-fits-all answer.
It highly depends on the hardware architecture, the specific pair style, and the cutoff distance for the force computation.
Figure~\ref{fig:pair} demonstrates the impact of differing combinations of neighbor list settings for the LJ potential for two different GPU architectures, NVIDIA H100 (red curves/squares) and AMD MI250X (blue curves/squares). 
Generally speaking, the more compute intensive a pair style is the more likely it is that half neighbor lists are the right choice.
Furthermore, on some architectures, such as NVIDIA GPUs, the atomic throughput is high enough that the overhead of atomics can be lower than the cost of the redundant computation performed in full neighbor lists.

The common pair style implementations also provide variants of the force kernel that leverage Kokkos hierarchical parallelism to expose the concurrency over the neighbors of each atom.
This is in contrast to the default approach of one work item per atom.
This can significantly improve performance for small problem sizes, where the number of atoms is not sufficient to saturate the hardware concurrency of modern GPUs.

The non-Kokkos implementation of pairwise forces does not have this common base implementation approach, leading to significant amounts of code duplication.
The Kokkos implementation avoids this; it is a unified source for the logic and implementation of the multiple execution policies described above.

One performance critical capability of Kokkos exploited in pairwise force computations is the transparent data layout adjustment of Kokkos \texttt{View}s.
To achieve good data access patterns on CPU architectures, the neighbor list for each atom must be contiguous in memory to enable caching, while the neighbor lists of consecutive atoms must be interleaved to achieve performance on GPU architectures.
Using 2D \texttt{View}s to implement the neighbor list achieves this data layout adjustment by default.
This effect was described as early as in the original Kokkos paper \cite{edwards2014} in Fig. 12. 

A separate issue affecting GPU throughput is thread divergence arising from each atom having differing numbers of neighbors and the conditional caused by the force cutoff check.
We will discuss approaches to addressing divergence in further detail later.


\subsection{ReaxFF}\label{sec:casestudies-reaxff}

In addition to the basic interatomic interactions described in section~\ref{sec:casestudies-pairwise} above, molecular dynamics codes also support potentials that explicitly model bonds between atoms.
The addition of bonded states allows the simulation of entirely new categories of materials, such as biomolecules, catalysts, and polymers.
LAMMPS contains a number of ways to model bonds in materials, and especially important amongst them is ReaxFF, or the Reactive Force Field \cite{van2001reaxff}. 
The term ``reactive'' in this context refers to the ability to not only model basic (static) bond parameters such as bond distances, angles, and energies, but also the dynamic formation and dissociation of bonds during a simulation.
This additional capability allows for the simulation of complex chemical processes and the study of reaction pathways and kinetics~\cite{senftle2016reaxff}.

The increase in chemical fidelity simultaneously increases computational demands. 
These benefits and costs have made ReaxFF a useful potential for LAMMPS benchmarking and a natural target for performance improvement efforts.
The initial CPU implementation of ReaxFF in LAMMPS is in the \texttt{REAXFF} package \cite{REAXFF}, which was extended with parallel OpenMP support within the \texttt{OPENMP} package \cite{REAXFF_OMP}. 
Later, the ReaxFF implementation in LAMMPS was ported to the Kokkos backend, maintaining host capabilities but extending the implementation to GPUs.
The performance improvements made to ReaxFF via the \texttt{KOKKOS} package provide a useful overview of several common patterns used in successful ports.
We briefly introduce each of these patterns below.


The first pattern is reducing thread divergence on the GPU.
The \texttt{KOKKOS} package typically uses a one-atom-per-thread model, where each thread has a serial loop over neighbors.
If the distance between an (atom, neighbor) pair of atoms is within the cutoff radius $r_\text{c}$ in one thread but not the other, the latter thread sits idle.
This effect gets exponentially worse for multi-body forces. This was solved with \textit{data pre-processing}. Work was split into two phases. There is a divergent pre-processing phase where the relatively inexpensive \textit{conditionals} are evaluated and a compressed table of interactions is constructed. The compressed interaction table then enters a fully convergent work phase.

The next step involved increasing the exposed parallelism, either as part of pre-processing as described above or in the form of \textit{hierarchical} parallelism as enabled by Kokkos.

The last step was unrelated to the programming model, but was important for \textit{exascale preparedness}.
This was proactively considering sources of 32-bit integer overflow seen in large simulations and replacing variables with 64-bit integers as appropriate. We discuss this in section~\ref{sec:reaxint}\footnote{See the following GitHub pull requests: \url{https://github.com/lammps/lammps/pull/4207} and \url{https://github.com/lammps/lammps/pull/4318}.}.

The following section will focus on four kernels to flesh out the above discussion.
First, to motivate work pre-processing to reduce divergence, we separately discuss the bond order neighbor list kernel and the four-body force kernel. 
Second, to motivate increasing parallelism to improve performance, we discuss the kernel that builds the sparse matrix in the charge equilibration step. 
Third, we briefly discuss fusing iterative sparse matrix solves to promote matrix reuse.
These kernels will be explored using a key LAMMPS ReaxFF benchmark, a short simulation of the molecular crystal Hexanitrostilbene (HNS).

\subsubsection{ReaxFF Optimization: Reducing Divergence via Pre-Processing Kernels}\label{sec:reaxpreprocessing}

The four-body force considers potentially bonded quads of atoms $i, j, k, l$.
Each work-item corresponds to one atom $i$ and contains a triply-nested loop over possible $j$, $k$, $l$. 
The quad of atoms contributes to the torsion force if $(i, j)$ are bonded, $(i, k)$ are bonded, and $(j, l)$ are bonded. 
There is also a constraint on the product of the bond orders. 
For HNS, in practice fewer than 5\% of possible quads satisfy each constraint, which leads to a high degree of divergence.

The four-body force calculation itself has a high computational intensity---multiple addition, multiplication, division, and transcendental function evaluations---and as such unused threads are wasted compute throughput. 
The solution here is to split the kernel into two divergent but relatively inexpensive pre-processing kernels and a fully convergent computation kernel.

The pre-processing phases determine the quads of indices $(i, j, k, l)$ that obey the aforementioned constraints and saves them to a Kokkos View of \texttt{int4} data types. 
The first pre-processing kernel counts the total number of quads and stores the per-atom-$i$ count, the Kokkos View is resized if necessary, and the second pre-processing kernel stores the quads. 
The View of quads is populated using a thread-safe global queue.
With this process, all quads for an atom $i$ are guaranteed to be contiguous.

The updated computation kernel is now fully convergent because it only acts on quads obeying the requisite constraints. 
In addition, we can further improve performance by parallelizing over \textit{quads} instead of atoms. 
Consecutive work-items will in general reuse some subset of $i$, $j$, $k$, and $l$ values because all quads for atom $i$ are contiguous.
This promotes cache reuse, boosting performance.

This discussion carries over exactly to the three-body force, except now we are only concerned with triplets of atoms and the rate of divergence is lower; nonetheless this approach is still beneficial. 
In practice we \textit{fuse} the pre-processing kernels for the three- and four-body forces.
This further promotes cache reuse.

\subsubsection{Exploiting hierarchical parallelism}

The charge equilibration method in ReaxFF proceeds in two stages:
\begin{enumerate}
    \item Construct a sparse matrix which encodes electrostatic interactions between pairs of atoms. The values at any (row, column) value correspond to interaction strengths between (atom, neighbor) pairs.
    \item Perform two Krylov solves. The atom charge distribution is a function of the solutions to these linear systems.
\end{enumerate}
Both stages are repeated on each molecular dynamics timestep.

The sparse matrix format is a modification of Compressed Sparse Row (CSR) where instead of the matrix values being densely packed, the matrix is ``over-allocated''; the allocated space for each row is the maximum number of neighbors as opposed to the exact number of neighbors within the cutoff radius. 
The matrix is described by \textit{four} data structures: a \textit{flat} array of non-zero values, the column offsets for each value, the offset array, \textit{and} an additional array that specifies the ``number of non-zero'' elements per row. 
This is required because the matrix is over-allocated.

This format makes the sparse matrix build less expensive.
There is an initial parallel scan over the number of neighbors in the \textit{full} neighbor list, independent of bond cutoffs. 
This determines the offset array. 
Next there is a kernel that computes the non-zero matrix elements and injects the values into the non-zero value array, along with specifying the number of non-zeroes and the column offsets. 
As implemented, this process carries over to Kokkos by using parallel for, reduction, and scan dispatches as appropriate.

The na\"ive porting of these kernels to Kokkos exposes computing and applying each row as one unit of independent work. 
This is efficient on the host but not on the device because, on the device, it leads to divergent memory access patterns in the matrix build and application. This is a known issue with CSR-style formats. 
The known solution is to expose additional parallelism over each row.

This exposure was accomplished with Kokkos team hierarchical parallelism.
Each Kokkos ``thread'' corresponds to one row.
Individual matrix elements in a row are assigned to ``vector'' lanes, which map to GPU threads.
Because per-row work is distributed over contiguous threads, convergent memory patterns are restored.
This improves the performance of counting the number of non-zero elements per row (hierarchical reduction) as well as computing and slotting the values of the non-zero elements (hierarchical scan) into the modified CSR matrix data structures. This optimization was found to be unprofitable on the CPU, so the code was bifurcated to allow the original version to run on CPUs, as described in \ref{sec:exe-control}.




\subsubsection{Kernel Fusion}\label{sec:reaxff-kernel-fusion}

The charge equilibration process is implemented as solving two linear systems to a fixed relative residual tolerance. The core kernel here is the sparse matrix-vector operation, which is bandwidth bound. The sparse matrix structure is the largest data structure. It is also identical between the two solves. By fusing the solves, performance improves by reusing the matrix load. We note this optimization was implemented by AMD for the Kokkos version\footnote{See \url{https://github.com/lammps/lammps/pull/3147}.} and was originally in the OpenMP version of ReaxFF \cite{REAXFF_OMP}.

\subsection{SNAP}\label{sec:casestudies-snap}

Significant effort has gone into optimizing both the SNAP machine learning algorithms and Kokkos implementation \cite{trott2014snap,gayatri2020rapid,nguyen2021billion,snapunify,LASA2024155011}. In particular, we note that Figure 6 of Ref.~\cite{LASA2024155011} provides a partial history of the performance of SNAP over time, and shows that our optimizations for NVIDIA GPUs also gave speedup on AMD MI250X hardware.
We describe some key points below.

In contrast to Lennard-Jones and ReaxFF, SNAP encodes the atomic neighborhood of a given atom $i$ in a more general hyperspherical decomposition. 
The hyperspherical basis on its own encodes two-body interactions. 
The decomposition for each (atom, neighbor) pair is encoded in Wigner U-matrices and summed.  
However, it is not the U-matrices but appropriate triple products thereof that are basis-independent, and can encode up to four-body interactions. 

The decomposition is formulated as
\begin{equation}
{\bf U}_j(i) = \sum_{r_{ik}<R} {\bf u}_j;~~~~~~{\bf u}_j = \mathcal{F}({\bf u}_{j-1/2}).\label{eq:snapU}
\end{equation}
The Wigner U-matrices are $\bf u_j$, which are a set of matrices of rank $2j+1$. The $j$ are positive integers and half-integers. The $\bf u_j$ are defined recursively; the right-hand side of equation~\ref{eq:snapU} denotes this as a recursion relation where $\mathcal{F}$ is a linear operator mapping adjacent elements of ${\bf u}_{j-1/2}$ to elements of ${\bf u}_j$. The sum of these matrices over all neighbors $k$ of an atom $i$ gives the full description of the atomic neighborhood, ${\bf U}_j$.

 These triple products are
\begin{equation}
B_{j_1, j_2, j} = {\bf U}_{j_1} \otimes^j_{j_1,j_2} {\bf U}_{j_2} : {\bf U}^*_j = {\bf Z}^j_{j_1,j_2} : {\bf U}^*_j,\label{eq:descriptorB}
\end{equation}
where $\otimes^j_{j_1,j_2}$ denotes an $\mathcal{O}(j^4)$ operation and ``$:$'' represents an element-wise scalar product of matrices ($\mathcal{O}(j^2)$). Group-theoretical symmetries allow us to constrain the computation to $0 \le j_2 \le j_1 \le j \le J$, significantly reducing the required work and storage.

The potential form of SNAP is a linear combination of these triple products. In the general case, this is of the form

\begin{equation}
E = \sum_{j, j_1, j_2} {\bf \beta}^j_{j_1,j_2} B_{j_1, j_2, j},\label{eq:snapE}
\end{equation}
where the ${\bf \beta}$ are trained coefficients.
Since the $j$ values are unbounded, this can be an infinite sum. In practice, it is truncated both as a statement of feasibility and as a statement of diminishing returns of \textit{accuracy} of the machine-learned model. A reasonable upper bound is $J_\text{max} = 4$.

An appropriately trained and truncated potential can encode ``any'' two-, three-, and four-body interaction to some given resolution. We again defer to~\cite{thompson2015spectral} for the more granular details.

The force is the derivative of the potential form. The derivative can be written as
\begin{equation}
\vec{F}_{i, k} = \sum_{j} {\bf Y_j} : \frac{\partial {\bf U}^*_j}{\partial {\bf \vec{r}}_k}; {\bf Y}_j = \sum_{j_1j_2} {\bf \beta}^j_{j_1,j_2} {\bf Z}^j_{j_1,j_2},\label{eq:snapF}
\end{equation}
with $\bf Y$ being the so-called \textit{adjoint} matrix.
This relatively simple form is made possible by symmetries of the ${\bf U}$ matrices and the triple product.
The $\bf Y$ matrices are the same size as the $\bf U$ matrices.

Computing the SNAP force reduces to four steps that have a one-to-one mapping with the software implementation of the SNAP force evaluation. 
The initial, non-Kokkos CPU implementation proceeds as follows. 
There is an outer loop over all atoms. 
Within this outer loop, there are four subroutines corresponding to four steps of evaluating the SNAP force.

\begin{enumerate}
   \item {\texttt{ComputeUi}}. Compute per-(atom, neighbor) pair Wigner $\bf u$ matrices, which are a representation of hyperspherical harmonics efficiently computed via a linear recursion relation, and accumulate them into the per-atom $\bf U$ matrices. This itself has an ``intermediate'' loop over neighbors and an innermost loop over quantum numbers with a serial dependency.
    \item \texttt{ComputeYi}: Compute per-atom $\bf Z$ matrices, which are a function of $\bf U$ matrices, and accumulate them into the per-atom adjoint $\bf Y$ matrices. This has an inner loop over quantum numbers of the $\bf Z$ matrices.
   \item \texttt{ComputeDuidrj}: Compute per-(atom, neighbor) pair derivatives of the Wigner $\bf u$ matrices, $\bf \vec{d}u$. This again has an ``intermediate'' loop over neighbors and an inner-most loop over quantum numbers with a serial dependency.
  \item \texttt{ComputeDeidrj}: Compute per-(atom, neighbor) pair force contributions $\bf \vec{F}$ by contracting $\bf Y$ and $\bf \vec{d}u$ over quantum numbers.
\end{enumerate}

Each one of these kernels requires staging space for intermediate data. 
Because this initial implementation was serial, it had a single data structure \textit{without} an atom index because data structures could be reused across outer loop iterations. 
The only ``non-trivial'' data layout questions were those of promoting cache reuse (discussed in more detail below).

We lost the ability to reuse intermediate data structures across individual atom calculations when we switched to a parallel implementation.
Each data structure needed a new ``atom index'' degree of freedom.
This required a careful consideration of data layouts because, in the most general case, data locality and cache reuse are not guaranteed \cite{gayatri2020rapid}.

Kokkos provides a natural framework for this performance-portable generalization through \texttt{View}s. 
A na\"ive porting was relatively straightforward. Next, Kokkos's native support for multi-dimensional parallel dispatches---enabling cache blocking and hierarchical parallelism---and its scratchpad memory abstraction permitted non-trivial optimizations to the SNAP implementation. 
Its support for algorithmic specialization enabled host/device code bifurcation. 
This was used for two kernels that benefited from the high arithmetic intensity permitted by GPUs but would be slower on the CPU. This is analogous to how, for the Lennard-Jones potential, it's advantageous to use a \textit{full} neighbor list even if it requires redundant work.

The implementation of SNAP demonstrates the flexibility of the Kokkos framework. 
Kokkos exposes the features necessary to perform most non-trivial optimizations in a performance portable way, and provides the ability to break out of single-source implementations when peak performance demands it.

\subsubsection{SNAP Data Structures and Kokkos Views}

The core data structures in SNAP are the sets of $\bf U$ and $\bf Y$ matrices stored as 16 byte complex doubles. 
As part of the transition to Kokkos we had to add an ``atom'' index. 
There are thus four degrees of freedom: the atom index $i$, the half-integer matrix index $j$, the matrix row index $m$, and the matrix column index $m'$. 
This is transformed to a two-dimensional regular data structure by flattening the $j, m, m'$ triplets into a single ``quantum number'' index. 
We use a $j$ slowest, $m'$ fastest convention to promote locality: rows and columns of matrices stay together.

There is a natural hierarchy of locality in these data structures. 
For each atom, the full set of $\bf U$ and $\bf Y$ are of size $\mathcal{O}(J^3)$, individual matrices ${\bf U}_j$ and ${\bf Y}_j$ of size $\mathcal{O}(J^2)$, and individual rows or columns thereof with size $\mathcal{O}(J)$. 
Performance is maximized by exploiting this locality in how we formulate algorithms and design data layouts

This data structure naturally translates to a Kokkos View. 
On the host, the quantum number index is fastest and atom index is slowest; this complements data locality as work is done one atom at a time. 
On the device, the atom index is fastest and the quantum number is slowest; this complements the benefits of memory coalescing on GPUs.
We use Kokkos multi-dimensional parallel dispatches to implement cache tiling to promote reuse.
This is described in more detail below.

\subsubsection{SNAP ComputeYi Implementation}\label{sec:snap-computeyi}

We first describe the second step in the SNAP force evaluation. There are two sources of parallelism. 
First, there is trivial parallelism over atoms. 
Second, there is parallelism over all $\mathcal{O}(J^5)$ components of ${\bf Z}^j_{j_1,j_2}$ (so long as we employ atomic operations). 
To promote data locality we flatten the quantum numbers for ${\bf Z}$, $j, j_1, j_2, m, m'$ into a linear index, with $j$ slowest, and $m'$ fastest. 

 The \texttt{ComputeYi} kernel requires computing each element of ${\bf Z}^j_{j_1, j_2}$, which is the computation of  weighted complex dot products of subsets of pairs of $\bf U$ matrices. This has a low computational intensity and trends memory bound.
By keeping temporal locality in computing $(j_1, j_2)$ pairs of ${\bf Z}$ components, we get cache reuse of ${\bf U}_{j_1}$ and ${\bf U}_{j_2}$. 
These can reside well in the lowest-level caches. 
Similarly, by preserving temporal locality in computing \textit{all} components of ${\bf Z}^j_{j_1, j_2}$; alternatively, locality in computing all of ${\bf Y}_j$, we can achieve cache reuse for the full set of $\bf U$ for a given atom.

On the CPU, we can achieve locality if (a) our data layout for ${\bf U}$, which is read, is quantum number fastest, atom number slowest and (b) our loop order is ``flat'' ${\bf Z}$ index fastest, atom slowest. 
This is achieved out-of-the-box with Kokkos Views.

While we can transpose the data layout on the GPU, we cannot na\"ively transpose the loop order and achieve locality. 
If, at fixed flattened ${\bf Z}$ index, we loop over all atom indices, the relevant components of $\bf U$ for the first atom falls out of lower-level caches after traversing \textit{all} atoms. 
The solution to this is a 3-d tiled traversal to get good data access patterns \textit{and} locality, as inspired by the Cabana framework~\cite{slattery2022cabana}.

The tiling pattern is $N_\text{atom}~\text{mod}~v$ fastest, flattened ${\bf Z}$ index middle, $N_\text{atom}~\text{div}~v$ slowest, where $v$ is a ``batch'' size. 
$v$ needs to be large enough to achieve well-behaved memory transactions (and work convergence) but small enough such that the dependent data for $v$ atoms times $\mathcal{O}(J^4)$ components of ${\bf U}$ reside well in caches.

The ideal value of $v$ is architecture-dependent, with a non-trivial sensitivity to the granularity of work and to cache sizes.
An intuitive lower bound is the warp (or equivalent) granularity, $v = 32$ on NVIDIA GPUs and $v = 64$ on AMD GPUs. 
The reasonable upper bound can be inferred by experiment but connects to L1 cache size. 
It was found in practice that the ideal values for $v$ are 32 on NVIDIA GPUs and 16 on Intel GPUs. 
Perhaps non-intuitively, the ideal value on AMD GPUs was seen to be $v = 32$.\footnote{These values are encoded at \url{https://github.com/lammps/lammps/blob/4246fab5/src/KOKKOS/kokkos_type.h\#L1379-L1387}} 
Kokkos enables this explicit experimentation and tuning which is important for maximizing performance.
A deeper discussion of the specific role of cache size can be found in section~\ref{sec:casestudies-cache}.

\subsubsection{ComputeUi and ComputeDuidrj}

The computation also exposes two levels of parallelism. 
There is trivial parallelism over atoms. 
If we permit thread atomic additions, we can also parallelize over neighbor atoms. 
We cannot (trivially) parallelize over quantum numbers because of the serial dependency in the recursion relation.

The initial implementation of \texttt{ComputeUi} and \texttt{ComputeDuidrj} was memory bound. 
Each discrete unit of work, (atom, neighbor) pairs, looped over each $j$ value and evaluated the hyperspherical harmonic recursion relation, staging the full set of ${\bf u}_j$ for each pair in memory. 
Once all ${\bf u}_j$ have been computed, they were sequentially reloaded and atomically accumulated into ${\bf U}_j$.
The recursion relation has a low arithmetic intensity making it memory bandwidth bound.

The ideal loop order on the CPU is neighbor index fastest, atom index slowest, in close analogy with \texttt{ComputeUi}.
The ideal data layout for ${\bf u}$ is quantum number (flattened $j, m, m'$) fastest to promote cache reuse in evaluating the recursion relation.
The ideal ``middle'' index and slow index is neighbor and atom index, respectively, preserving data locality in atomic accumulation into ${\bf U}$.
Similar to \texttt{ComputeYi}, we have transposed data layouts on the GPU and implement a tiled data traversal for cache locality.

Further optimizations were possible on the GPU.
This is based on the critical observation that \textit{recursive polynomial evaluation is inherently compute bound}. 
The only inputs are the base case and all components of $\bf u$ follow from that.

Realizing this optimization required four separate and simultaneous optimizations. 
First, we re-wrote the recursive polynomial evaluation as a hybrid depth/breadth evaluation, depth first in rows of ${\bf u}_j$. 
Relative to the initial breadth-first form, this reduces intermediate state overheads from $\mathcal{O}(j^2)$ to $\mathcal{O}(j)$. 
Second, we atomically accumulated components of ${\bf u}_j$ directly into ${\bf U}_j$, eliminating the additional round-trip to memory. 
Third, we \textit{explicitly} cached intermediate values in Kokkos scratchpad memory, enforcing locality. 
Last, we implemented a redundant work model which recomputes some rows of ${\bf u}_{j' < j}$ while computing unique rows of ${\bf u}_j$. 
This let us completely eliminate all staging in global memory.

This approach carries over directly to \texttt{ComputeDuidrj} except we now stage the matrices ${\bf U}$ and the derivative ${\bf \vec{d}u}$. 
In place of the atomic accumulation into ${\bf U}$ we can inline the accumulation into the force $\bf \vec{F}$---which is a kernel fusion with \texttt{ComputeDeidrj}. We denote this fused kernel \texttt{ComputeFusedDeidrj}.

This approach lays bare the relative speeds and feeds for the CPU versus the GPU. 
This approach is advantageous on the GPU because the redundant work is more than amortized by the savings from manually managed data locality. 
There is not the same benefit on the CPU due to the different balance of computation and memory throughput; the redundant work defeats potential performance gains.

\subsubsection{Work Batching and Instruction Level Parallelism}

The unifying feature of our last set of optimizations is \textit{work batching}, where each thread handles multiple units of \textit{independent but identical work}. Work batching can be used to shift limiters. It is also a way to expose \textit{instruction level parallelism} (ILP). This is because the compiler is ``free'' to interleave operations on independent work, hiding serial dependencies, and possibly improving throughput.

We have already seen work batching in ReaxFF when the the two sparse linear system solves were fused, described in section~\ref{sec:reaxff-kernel-fusion}. In the sparse matrix-vector operation, the units of independent work are loading components of the right-hand-side vector, multiplying by the \textit{same} matrix element, and accumulating into independent output vectors. Our limiter is loading the matrix elements. We can simultaneously hide the latency of loading both vectors behind the latency of loading the matrix elements. This is the origin of the performance improvement.

This optimization is specific to sparse matrix-vector operations. It was informed by quantitatively identifying the kernel performance limiter and considering if and how we could use work batching to improve end-to-end performance.

We followed this same roadmap for implementing batching in the SNAP kernel. Our initial optimization efforts focused on NVIDIA H100. Limiters were identified using NVIDIA Nsight Compute. However, the idea of work batching to shift limiters and improve ILP is not specific to any one architecture. By introducing the batch size as a compile-time tunable parameter, we could find performance improvements on AMD MI300A with otherwise identical code.

\begin{table}[tbp]
  \centering
  \caption{
  Relative performance uplift from work-batching optimizations on NVIDIA H100 and AMD MI300A GPUs for the top three kernels. We include the explicit batch factors for ComputeUi and ComputeYi. For ComputeFusedDeidrj we simply fused all three directions.}
  \label{tab:gpu-batching-optimizations}
\begin{tabular}{|l|c|c|}
    \hline
    Kernel & MI300A Speed-up & H100 Speed-up \\ \hline
    ComputeUi & $1.75\times$ (batch 2) & $2.23\times$ (batch 4) \\
    ComputeYi & $1.04\times$ (batch 4) & $1.54\times$ (batch 4) \\
    ComputeFusedDeidrj & $1.74\times$ & $1.49\times$ \\
    \hline
  \end{tabular}
\end{table}





The \texttt{ComputeUi} kernel was limited by double precision floating point addition from the accumulation of the components of ${\bf u}_j$ for each neighbor atom $k$ into ${\bf U}_j$. 
We can reduce the number of atomic additions if we have each thread handle one atom but multiple neighbors. 
Before each atomic addition, the thread can perform the sum over neighbors \textit{locally} before performing the atomic addition into ${\bf U}_j$. 
This shifts the limiter away from atomic addition, improving performance.

There is an additional benefit from this optimization. 
Multiple instances of the recursive polynomial evaluation, one for each local neighbor $k$, are now performed within each thread.
This exposes a multi-fold amount of \textit{independent} floating point work. 
This improves the ability of the compiler to interleave independent work, improving floating point throughput.

The \texttt{ComputeYi} kernel was limited by L1 cache throughput. This had two sources. One is loads of ${\bf U}_j$ where each thread handled one atom. The other is look-up tables which were constant across all threads in a warp (or equivalent on non-CUDA architectures), that is, independent of the atom index. There is a serial data dependency on the data in the look-up table and as such it could not be hidden.

We can reduce the number of accesses to these look-up tables relative to loads of ${\bf U}_j$ if we have each thread handle multiple atoms. 
This batching does not change the limiter, L1 cache throughput, but reduces the total number of latency-driven transactions, improving performance.

Each instance of the \texttt{ComputeFusedDeidrj} kernel was already relatively well-optimized because, between computing ${\bf U}_j$ and ${\bf dU}_j$, there were independent floating point calculations. 
The key observation is there was redundant work between each separate instance of the kernel, one for each direction. 
The redundant work was re-computing ${\bf U}_j$ and re-loading ${\bf Y}_j$ each time.

By fusing the three kernels together, we reduced the amount of redundant floating point computation and eliminated repeated loads of ${\bf Y}_j$. 
This latter point was, in practice, a source of stalls due to serial data dependencies.

We list the performance improvements for all three kernels in table~\ref{tab:gpu-batching-optimizations}. As noted before, these optimizations were driven by a quantitative analysis of limiters on NVIDIA H100. We could then \textit{empirically} re-tune the batching factors on AMD MI300A and see if we achieved a speed-up. In the ``worst case'', a batching factor of $1$ would restore the original performance.

On AMD MI300A we did find performance improvements with non-trivial batching factors. This speaks to the broad applicability of improving instruction-level parallelism. In the case of \texttt{ComputeUi} and \texttt{ComputeFusedDeidrj}, our work batching drove the known limiters towards double-precision compute, which is a known strength of modern GPUs.

\subsection{Cache Size Performance Impact}\label{sec:casestudies-cache}

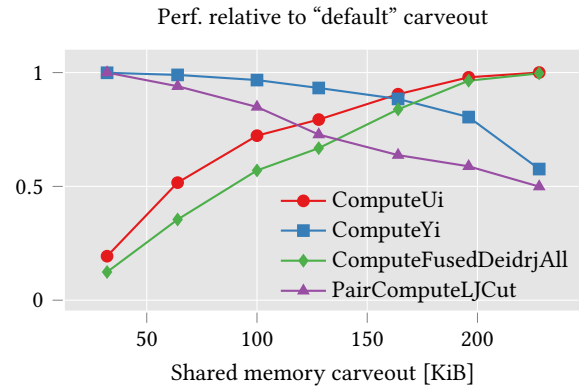
\begin{figure}[tbp]
    \centering
\if\regenfigs1
\begin{tikzpicture}
\begin{axis}[
    thick,
    legend style={at={(1.0,0.25)},anchor=east},
    xlabel={Shared memory carveout [KiB]},
    title={Perf. relative to ``default'' carveout},
    height=2in,
    ymin=-0.05,
    width=\linewidth,
]
    \addplot+[mark=*] table[y index=1] {results/cache-carveout-h100-hbm3/carveout_data_h100_hbm3_new.txt};
    \addlegendentry{ComputeUi};
    \addplot+[mark=square*] table[y index=2] {results/cache-carveout-h100-hbm3/carveout_data_h100_hbm3_new.txt};
    \addlegendentry{ComputeYi};
    \addplot+[mark=diamond*] table[y index=3] {results/cache-carveout-h100-hbm3/carveout_data_h100_hbm3_new.txt};
    \addlegendentry{ComputeFusedDeidrjAll};
    \addplot+[mark=triangle*] table[y index=4] {results/cache-carveout-h100-hbm3/carveout_data_h100_hbm3_new.txt};
    \addlegendentry{PairComputeLJCut};
\end{axis}
\end{tikzpicture}
\else
\includegraphics{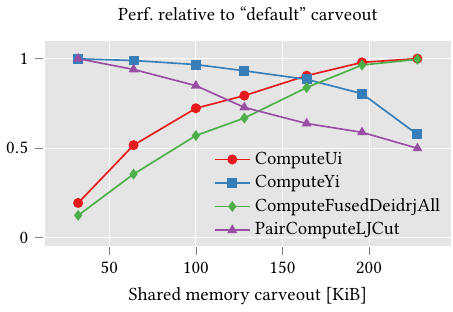}
\fi
    \caption{Performance of the \texttt{ComputeUi}, \texttt{ComputeYi}, and \texttt{ComputeFusedDeidrjAll} kernels in SNAP and the pairwise force kernel \texttt{PairComputeLJCut} in Lennard-Jones as a function of the shared memory carveout on NVIDIA H100-HBM3-SXM. The performance is normalized against the ``default'' value selected at runtime. All runs were at 1,024,000 atoms.}
    \label{fig:cache_carveout}
\end{figure}
 

It can be nontrivial to undertake cross-architecture investigations of performance. 
The Kokkos library eases this challenge considerably by offering a single source of flexible code with which to do controlled comparisons. 
One application of this capability that we demonstrate is investigating the sizes of L1 cache and shared memory capacities, which have known differences between GPU architectures. 

Generally speaking, GPUs have two sets of L1-like cache: a hardware and a software managed one.
The latter is referred to as \textit{shared memory} in CUDA, \textit{local data share} in HIP, and \textit{local memory} in SYCL.

The characteristics of these caches are one of the major differences of the GPU architectures used on exascale class systems (see table~\ref{tab:gpu-specs}).
The sizes of the L1 cache and shared memory are fixed on AMD MI250X and MI300A, as well as on Intel PVC, because they are discrete units. On the other hand, modern NVIDIA GPUs have a unified cache where the L1 and shared memory capacity can be dynamically shifted.

The ability to shift between L1 and shared memory on NVIDIA GPUs enables us to design an experiment to investigate the impact of L1 or shared memory size on kernel performance leveraging what CUDA calls the shared memory \textit{carveout}.
This is the percentage of the unified cache reserved for shared memory. 
With kernels that do not use shared memory but benefit from L1 cache, we expect performance to decrease with an increasing ratio of the combined cache being reserved for shared memory.
For kernels which use shared memory and do not rely on automatic caching in L1 the opposite is expected, and for ones that leverage both L1 and shared memory the best performance should be achieved with some intermediate carveout value.

Indeed Kokkos has a built-in heuristic to optimize the carveout value depending on the characteristics of the kernel launched.
For this study, we overwrote that heuristic and simply forced a specific carveout value to see its impact on four top kernels: the primary force computation kernel for the Lennard Jones potential \texttt{PairComputeLJCut}, and the three most time consuming kernels in SNAP, \texttt{ComputeUi}, \texttt{ComputeYi}, and \texttt{ComputeFusedDeidrj}. Kernel runtimes were measured using NVIDIA Nsight Systems. We also looked at the top four kernels for ReaxFF but did not see a significant dependence on the cache carveout (under 10\%).

The effect of manually setting the cache carveout on NVIDIA H100-HBM3-SXM is shown in figure~\ref{fig:cache_carveout}. 
The x-axis is the shared memory carveout and the y-axis is the performance of each kernel \textit{normalized by} the performance from running at the default carveout.

The force kernel of Lennard Jones does not use shared memory, but does see significant benefits from large L1 cache sizes.
At the maximum carveout for shared memory, which leaves only 32kB for L1, performance drops by about 50\%; alternatively, increasing the L1 cache size from 32kB to 224kB increases performance by 85\%.
The \texttt{ComputeYi} kernel shows a similar behavior; in this case the difference comes from reuse of $\mathbf{U}_j$ matrices instead of neighbor atom coordinates.

On the other hand \texttt{ComputeUi} and \texttt{ComputeFusedDeidrj} have their highest performance at the maximum shared memory carveout and performance drops with more of the combined resource dedicated to L1 cache. The scaling is nearly linear because occupancy is proportional to shared memory utilization.

Given these results, it is reasonable to assume that a significant fraction of the performance difference between otherwise similarly capable NVIDIA GPUs and AMD GPUs (specifically A100 vs. MI250x and H100 vs. MI300A) can be attributed to the cache size difference as well as the ability to tune the ratio of L1 to shared memory size on a per kernel basis.


\section{Performance Characteristics Across Architectures}\label{sec:results}

Our optimized implementations demonstrate several key insights about performance optimization strategies for molecular dynamics simulations on modern hardware architectures.

\begin{figure}[tbp]
\if\regenfigs1
\begin{tikzpicture}
\begin{axis}[
    thick,
    xmode=log,
    legend style={at={(0.9,0.4)},anchor=east},
    xlabel={\# atoms},
    ylabel={\% of peak performance},
    height=2in,
    width=\linewidth,
]
    \addplot+[mark=*] table[y expr={100*\thisrowno{1}/1335.066624}] {results/satlj_h100.txt};
    \addlegendentry{LJ};
    \addplot+[mark=square*] table[y expr={100*\thisrowno{1}/4.067584}] {results/satsnap_h100.txt};
    \addlegendentry{SNAP};
    \addplot+[mark=diamond*] table[y expr={100*\thisrowno{1}/11.378491}] {results/satreax_h100.txt};
    \addlegendentry{ReaxFF};
\end{axis}
\end{tikzpicture}
\else
\includegraphics{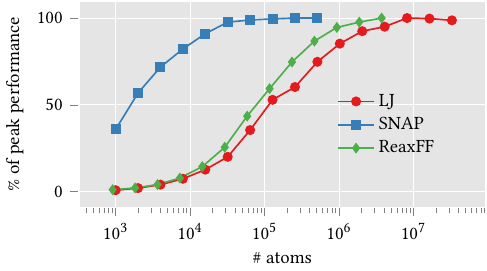}
\fi
\caption{Saturation of the normalized performance (atom-steps/s) on one NVIDIA H100 GPU for the three case studies. The much greater available parallelism of the ML-based SNAP potential allows it to run more efficiently for smaller system sizes. Note that ReaxFF ran out of HBM before reaching full saturation.}\label{fig:perfsat}
\end{figure}
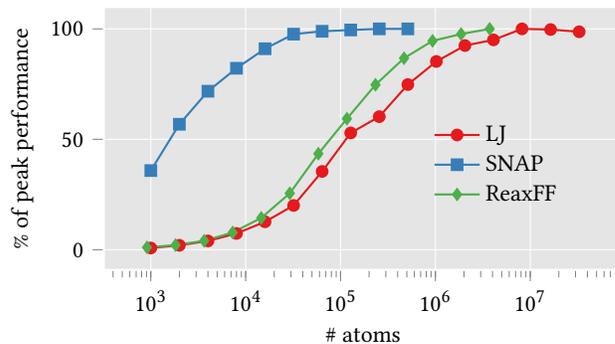

\subsection{Single GPU Comparison}\label{sec:results-arch}
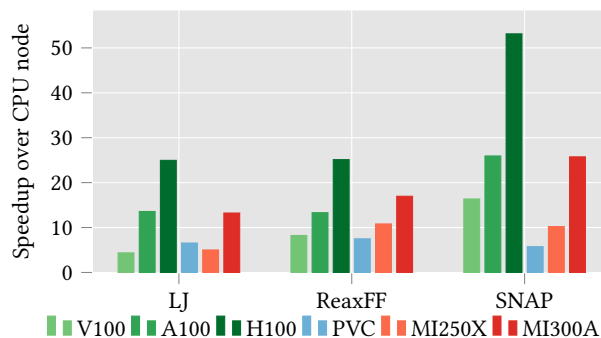
\begin{figure}[tbp]
\if\regenfigs1
\begin{tikzpicture}
\begin{axis}[
    ylabel={Speedup over CPU node},
    symbolic x coords={LJ,ReaxFF,SNAP},
    enlarge x limits=0.25,
    legend style={at={(0.45,-0.14)},
                  anchor=north,legend columns=-1},
    ybar,
    every axis plot/.append style={fill},
    ymin=-0.1,
    bar width=6pt,
    xtick=data,
    height=2in,
    width=\linewidth,
    ytick distance = 10,
]
    \pgfplotsset{cycle list/Greens-6}
    \pgfplotsset{cycle list shift=3}
    \addplot+ table[y index=2] {results/bench.txt};
    \addlegendentry{V100};
    \addplot+ table[y index=3] {results/bench.txt};
    \addlegendentry{A100};
    \addplot+ table[y index=4] {results/bench.txt};
    \addlegendentry{H100};
    \pgfplotsset{cycle list/Blues-5}
    \pgfplotsset{cycle list shift=4}
    \addplot+ table[y index=8] {results/bench.txt};
    \addlegendentry{PVC};
    \pgfplotsset{cycle list/Reds-5}
    \pgfplotsset{cycle list shift=3}
    \addplot+ table[y index=5] {results/bench.txt};
    \addlegendentry{MI250X};
    \addplot+ table[y index=6] {results/bench.txt};
    \addlegendentry{MI300A};
\end{axis}
\end{tikzpicture}
\else
\includegraphics{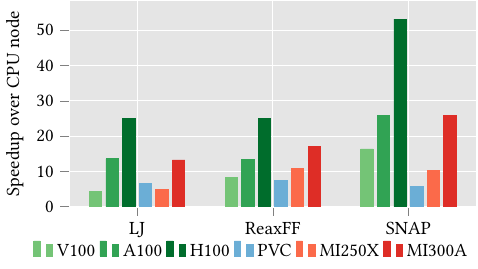}
\fi

\caption{Performance comparison of different generations of GPU hardware from NVIDIA, AMD and Intel GPUs for the three performance case studies (LJ: 16M atoms, ReaxFF: 465K atoms, SNAP: 64K atoms). The performance is normalized by the performance on a 36-core Skylake CPU node, using the base non-Kokkos LAMMPS code and MPI. Note that the Intel PVC and AMD MI250X performance was measured on one stack and GCD respectively, i.e., ``half the GPU''.}\label{fig:arch-perf}
\end{figure}


\definecolor{eos}{RGB}{35,139,69}
\definecolor{alps}{RGB}{35,139,69}
\definecolor{aurora}{RGB}{107,174,214}
\definecolor{frontier}{RGB}{239,59,44}
\definecolor{elcap}{RGB}{253,141,60}
\begin{figure*}[tbp]
\if\regenfigs1
\begin{tikzpicture}[overlay=false,>=latex]
    \begin{groupplot}[
            /tikz/overlay=false,
            /tikz/thick,
            group style={
                group size=3 by 1,
                vertical sep=0.2cm,
                horizontal sep=0.9cm,
                xlabels at=edge bottom,
                ylabels at=edge left,
                xticklabels at=edge bottom,
            },
            width=\linewidth/2,
            ylabel near ticks,
            xlabel={\# GPU nodes},
            xmode=log,
            ymode=log,
            log basis x=10,
            xmin=0.5, xmax=11000,
            clip=false,
            title style={yshift=-0.5cm, xshift=-2.1cm,anchor=west},
            ylabel={timesteps/s},
            height=2.1in,
            width=2.5in,
        ]
            
            \nextgroupplot[
                ymin=0.2, ymax=5050,
                legend style={at={(0.7,0.3)},anchor=east},
                legend columns=1,
                title={a) LJ},
            ]
            \addplot+[mark=*,frontier]  table[y index=2] {results/strongscaling/results_frontier/lj_1e7.txt};
            \addplot+[mark=triangle*,aurora]  table[y index=2] {results/strongscaling/results_aurora/lj_1e7.txt};
            
            \addplot+[mark=diamond*,elcap]  table[y expr={\thisrowno{1}*0.8192}] {results/strongscaling/results_elcap/lj_1e7.txt};
            \addplot+[mark=diamond*,elcap]  table[y expr={\thisrowno{1}*1.05}] {results/strongscaling/results_elcap/lj_1e9.txt};
            \addplot+[mark=diamond*,elcap]  table[y expr={\thisrowno{1}*1.342}] {results/strongscaling/results_elcap/lj_1e11.txt};
            
            \addplot+[mark=*,frontier]  table[y index=2] {results/strongscaling/results_frontier/lj_1e9.txt};
            \addplot+[mark=*,frontier]  table[y index=2] {results/strongscaling/results_frontier/lj_1e11.txt};
            
            
            \addplot+[mark=square*,alps]  table[y expr={\thisrowno{2}*0.8192}] {results/strongscaling/results_alps/lj-8M-4gpus.txt};
            \addplot+[mark=square*,alps]  table[y expr={\thisrowno{2}*1.05}] {results/strongscaling/results_alps/lj-1B-4gpus.txt};
            \addplot+[mark=square*,alps]  table[y expr={\thisrowno{2}*1.342}] {results/strongscaling/results_alps/lj-128B-4gpus.txt};

            \addplot+[mark=triangle*,aurora]  table[y index=2] {results/strongscaling/results_aurora/lj_1e9.txt};
            \addplot+[mark=triangle*,aurora]  table[y index=2] {results/strongscaling/results_aurora/lj_1e11.txt};
            
            \draw (axis cs:1,175) node[anchor=north] {\(10^7\)};
            \draw (axis cs:2,2) node[anchor=135] {\(10^9\) atoms};
            \draw (axis cs:128,1) node[anchor=north] {\(10^{11}\)};
            
            \nextgroupplot[
                ymin=0.4, ymax=205,
                legend style={at={(0.5,1.0)},anchor=south},
                legend columns=9,
                title={b) ReaxFF},
            ]
            \addplot+[mark=square*,alps]  table[y expr={\thisrowno{2}*0.9339}] {results/strongscaling/results_alps/reaxff-1M-4gpus.txt};
            \addlegendentry{Alps (4\(\times\)GH200)};
            \addlegendentry{Frontier (4\(\times\)MI250X)};
            \addplot+[mark=diamond*,elcap]  table[y expr={\thisrowno{1}*0.9339}] {results/strongscaling/results_elcap/hns_1e6.txt};
            \addlegendentry{El Capitan (4\(\times\)MI300A)};
            \addplot+[mark=triangle*,aurora]  table[y index=2] {results/strongscaling/results_aurora/hns_1e6.txt};
            \addlegendentry{Aurora (6\(\times\)PVC)};
            
            \addplot+[mark=*,frontier]  table[y index=2] {results/strongscaling/results_frontier/hns_1e8.txt};
            \addplot+[mark=*,frontier]  table[y index=2] {results/strongscaling/results_frontier/hns_1e10.txt};
            
            \addplot+[mark=square*,alps]  table[y expr={\thisrowno{2}*1.195}] {results/strongscaling/results_alps/reaxff-128M-4gpus.txt};
            \addplot+[mark=square*,alps]  table[y expr={\thisrowno{2}*1.53}] {results/strongscaling/results_alps/reaxff-16B-4gpus.txt};

            \addplot+[mark=diamond*,elcap]  table[y expr={\thisrowno{1}*1.195}] {results/strongscaling/results_elcap/hns_1e8.txt};
            \addplot+[mark=diamond*,elcap]  table[y expr={\thisrowno{1}*1.53}] {results/strongscaling/results_elcap/hns_1e10.txt};
            
            \addplot+[mark=triangle*,aurora]  table[y index=2] {results/strongscaling/results_aurora/hns_1e8.txt};
            \addplot+[mark=triangle*,aurora]  table[y index=2] {results/strongscaling/results_aurora/hns_1e10.txt};
            
            \draw (axis cs:1,18) node[anchor=north] {\(10^6\)};
            \draw (axis cs:4,0.96) node[anchor=north] {\(10^8\)};
            \draw (axis cs:512,1.24) node[anchor=north] {\(10^{10}\)};
            
            \nextgroupplot[
                ymin=0.0101, ymax=2050,
                legend style={at={(1.0,0.2)},anchor=east},
                legend columns=1,
                title={c) SNAP},
            ]
            \addplot+[mark=*,frontier]  table[y index=2] {results/strongscaling/results_frontier/snap_1e6.txt};
            
            \addplot+[mark=*,frontier]  table[y index=2] {results/strongscaling/results_frontier/snap_1e8.txt};
            \addplot+[mark=*,frontier]  table[y index=2] {results/strongscaling/results_frontier/snap_1e10.txt};
             

            \addplot+[mark=square*,alps]  table[y expr={\thisrowno{2}*1.024}] {results/strongscaling/results_alps/snap-1M-4gpus.txt};
            \addplot+[mark=square*,alps]  table[y expr={\thisrowno{2}*1.31}] {results/strongscaling/results_alps/snap-128M-4gpus.txt};
            \addplot+[mark=square*,alps]  table[y expr={\thisrowno{2}*1.68}] {results/strongscaling/results_alps/snap-16B-4gpus.txt};

            \addplot+[mark=triangle*,aurora]  table[y index=2] {results/strongscaling/results_aurora/snap_1e6.txt};
            \addplot+[mark=triangle*,aurora]  table[y index=2] {results/strongscaling/results_aurora/snap_1e8.txt};
            \addplot+[mark=triangle*,aurora]  table[y index=2] {results/strongscaling/results_aurora/snap_1e10.txt};
            
            \addplot+[mark=diamond*,elcap]  table[y expr={\thisrowno{1}*1.024}] {results/strongscaling/results_elcap/snap_1e6.txt};
            \addplot+[mark=diamond*,elcap]  table[y expr={\thisrowno{1}*1.31}] {results/strongscaling/results_elcap/snap_1e8.txt};
            \addplot+[mark=diamond*,elcap]  table[y expr={\thisrowno{1}*1.68}] {results/strongscaling/results_elcap/snap_1e10.txt};
            
            \draw (axis cs:1,9.3) node[anchor=north] {\(10^6\)};
            \draw (axis cs:1,0.096) node[anchor=north] {\(10^8\)};
            \draw (axis cs:32,0.031) node[anchor=north] {\(10^{10}\)};
    \end{groupplot}
\end{tikzpicture}
\else
\includegraphics{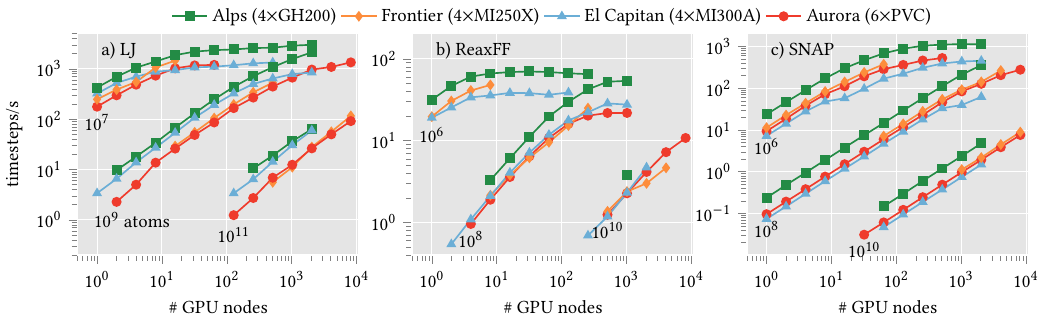}
\fi
\caption{Strong scaling results for LAMMPS on different exascale architectures. The total number of atoms is given by the label below each group of results. In cases of minor atom count discrepancies, the speed was rescaled assuming linear scaling. We see that SNAP, with its faster saturation (figure~\ref{fig:perfsat}) and higher computational expense delivers the best scaling, while ReaxFF with its lack of saturation plateau struggles to reach even 100 timesteps/s. Overall, we observe excellent strong scaling, even out to 8192 nodes on Frontier and El Capitan.}\label{fig:strongscaling}
\end{figure*}
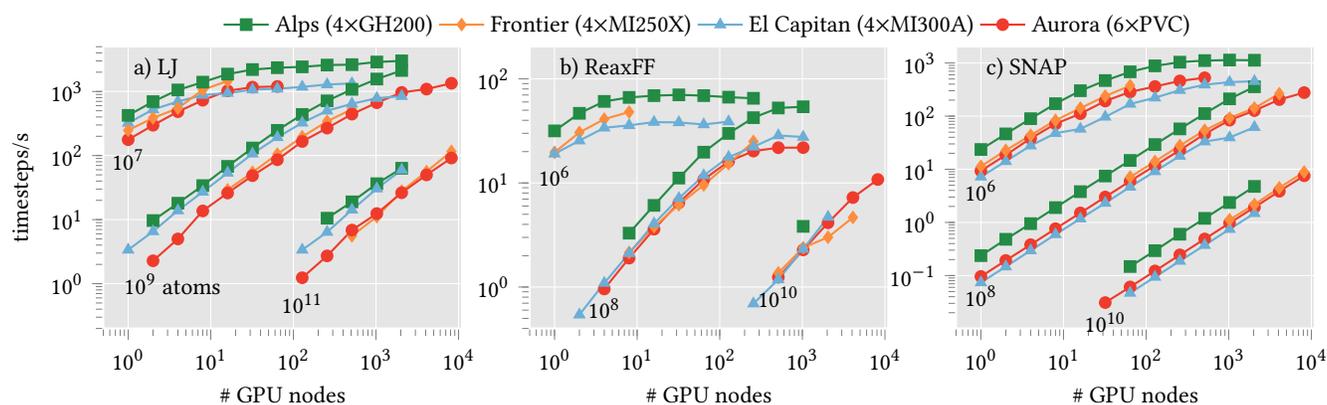

As discussed previously, the performance characteristics of our implementations vary significantly across different GPU architectures due to hardware-specific features such as cache sizes and thread atomics performance.
One of the most critical performance characteristics of GPUs is the large amount of parallelism a software has to expose to saturate a GPU, which now exceed 200,000 simultaneously active threads.

In figure~\ref{fig:perfsat}, we plot performance as a function of number of atoms on a single H100 GPU for the three case studies.
``Peak performance'' is the point at which short-range molecular dynamics simulations have saturated asymptotic linear scaling in the number of atoms (or run out of HBM).
At lower atom counts, hardware-induced thread starvation (not enough parallelism) and latency effects, such as kernel launch overheads, reduce the achievable performance.

We see that LJ and ReaxFF saturate at a similar point. 
This trend reflects similar latencies and similar degrees of exposed parallelism.
On the other hand, SNAP saturates at much lower atom counts, since the primary compute kernels expose several degrees of parallelism beyond just particle count as described in section~\ref{sec:casestudies-snap}.

In figure~\ref{fig:arch-perf}, we show results from a single GPU/GCD/stack of the three case studies on different hardware for a fixed atom size.
While the relative performance of the different architectures does generally follow the generic hardware specifics listed in table~\ref{tab:gpu-specs}, it is not simply determined by scaling with bandwidth or flop rate.
This is not unexpected since molecular dynamics kernels are cache sensitive, have unstructured data access and are not generally compute limited.
We suspect that a significant fraction of the difference is explained by the cache size differences. Modern NVIDIA GPUs have significantly larger L1 cache which also can be dynamically leveraged for software managed scratch.
We also believe that this difference explains the relatively large jumps between V100 and A100 which exceed the raw performance improvements in bandwidth and flop rate of the GPUs.

\subsection{Exascale System Scalability}\label{sec:results-scale}

In figure~\ref{fig:strongscaling} we look at the strong scaling of LAMMPS across all current exascale machines for the three case studies, scaling up to 8192 nodes on OLCF's Frontier and NNSA's El Capitan, as well as 2048 nodes on ALCF's Aurora. We additionally scale up to 2048 nodes of CSCS's Alps supercomputer and 256 nodes of a 4 GPUs per node on NVIDIA's Eos DGX Superpod; we intentionally run on four GPUs per Eos node to mimic the configurations of the largest NVIDIA-based supercomputers.\footnote{We defer presenting the results from Eos to appendix~\ref{sec:eos-alps} for the purpose of clarity and conciseness; the curves for Alps and Eos lie largely on top of each other due to strong similarities between H100 and GH200, as well as the comparable network bandwidths between NDR 400 and Slingshot-11. This is shown in figure~\ref{fig:eosalps}.}

Overall, we see excellent strong scaling performance across all machines and benchmarks, and LAMMPS achieves approximately 1000 timesteps/s for any problem size for LJ and SNAP provided enough nodes are available.
SNAP scales particularly well, with its higher computational expense hiding launch latencies and communication, as well as its lower atom count required for GPU saturation (figure~\ref{fig:perfsat}).
Due to its lack of a saturation plateau in figure~\ref{fig:perfsat}, ReaxFF shows the poorest scaling, as any node count greater than the smallest required to fit the problem size in memory immediately reduces the efficiency and prevents linear scaling.
As a result, no machine is able to exceed 100 timesteps/s for any system size.
Finally, we observe that the relative performance of the different machines is consistent with that of the single GPUs shown in figure~\ref{fig:arch-perf}.

\section{Conclusion}\label{sec:conclusion}

In this paper we discussed the design, optimization strategy, and performance of the LAMMPS KOKKOS package.
This package provides a single-source implementation of LAMMPS's capabilities that is performance-portable across CPUs and multiple vendors' GPUs via the Kokkos programming model.
While that performance portability is critical to be able to support various different architectures without needing to maintain separate code bases, optimization of the implemented algorithms is also essential to achieve good performance.
We demonstrated that Kokkos provides the necessary abstractions to leverage the hierarchical nature of parallelism on GPUs, expose hardware capabilities such as software managed caches, and facilitate the optimization of data access patterns.
These abstractions enabled non-trivial algorithmic improvements that give multiplicative performance improvements on top of generational hardware improvements.

In section~\ref{sec:casestudies} we described several optimizations across three molecular dynamics benchmarks.
These optimizations are motivated by GPU design principles: manage memory access patterns, consider arithmetic intensity, maintain code convergence, and expose maximal concurrency.

In section~\ref{sec:results-arch} we showed that, by following these principles, we extract performance across all GPU architectures: multiple generations of NVIDIA GPUs, as well as across GPUs from AMD and Intel.
This paper considered a source of performance deviation across architectures of similar capability: cache design.
Using cache carveout controls on NVIDIA H100 GPUs, we demonstrated performance drops of 20\% to 60\% across four top kernels when we matched the L1 cache or shared memory capacity, as appropriate, to that of AMD MI300A.
These deviations are consistent with the performance differences between these GPUs.

In section~\ref{sec:results-scale} we showed that performance of the LAMMPS KOKKOS package on modern supercomputers, including three exascale machines, achieves excellent scaling for the three benchmarks.
Generally, relative performance was dominated by single GPU performance with network effects subleading.
These investigations would not have been possible were it not for the Kokkos single-source performance portable framework and its integration into the LAMMPS software package for molecular dynamics.

\begin{acks}
We thank Christopher Knight at ALCF for invaluable help with using Aurora.
We thank Peter Messmer and Christelle Piechurski at NVIDIA, as well as Maria Grazia Giuffreda and Rocco Meli at CSCS, for their ongoing support with Alps runs at scale.
We also thank the many contributors to LAMMPS-KOKKOS, past and present, in particular Ray Shan, Rahul Gayatri, Trung Dac Nguyen, Dan Ibanez, Mitch Murphy, Richard Berger, Nick Curtis, Charlie Sievert, Sikandar Mashaya, Axel Kohlmeyer, and many others listed on the Authors
page \url{https://www.lammps.org/authors.html} on the LAMMPS website.
Sandia National Laboratories is a multimission laboratory managed and operated by National Technology and Engineering Solutions of Sandia, LLC., a wholly owned subsidiary of Honeywell International, Inc., for the U.S. Department of Energy's National Nuclear Security Administration under contract DE-NA-0003525.
This research was supported by the U.S. DOE Office of Science-Advanced Scientific Computing Research Program, under Contract No. DE-AC02-06CH11357 and used resources of the Argonne Leadership Computing Facility, a U.S. Department of Energy (DOE) Office of Science user facility at Argonne National Laboratory and is based on research, the Oak Ridge Leadership Computing Facility at the Oak Ridge National Laboratory, which is supported by the Office of Science of the U.S. Department of Energy under Contract No. DE-AC05-00OR22725, and the National Energy Research Scientific Computing Center (NERSC), a Department of Energy User Facility using NERSC award FES-ERCAP0034313.
\end{acks}

\bibliographystyle{ACM-Reference-Format}
\bibliography{refs}

\labelformat{section}{appendix~#1}
\appendix


\section{Enabling machine-learning-based interatomic potentials}
In addition to advancements in hardware, simulation methodology itself is rapidly evolving.
Over the past two decades, and particularly the last few years, one of the main trends in LAMMPS has been a shift away from empirical potentials (such as LJ and ReaxFF) towards machine-learning-based potentials, such as SNAP (the Spectral Neighbor Analysis Potential) \cite{thompson2015spectral} described in section~\ref{sec:casestudies-snap}.
These methods increase the quantitatively predictive power of the simulations through their foundation of using quantum mechanical calculations to provide atomic forces and energies as training data.
At the same time, they are also vastly more computationally expensive than traditional empirical potentials \cite{Plimpton2012}, making the efficient use of modern hardware such as GPUs critical.

LAMMPS and its KOKKOS package facilitate different approaches for integrating machine-learning-based potentials.
The first and typically most performant strategy is that taken by SNAP, FLARE~\cite{flaresummit}, ACE~\cite{ace}, and others, where the entire method is reimplemented using the Kokkos library.
From a LAMMPS development and compilation perspective, these potentials add minimal external dependencies and behave as any other potential. 
The downside to this approach is the associated labor of the manual implementation of the potential, especially its derivatives for force and virial calculation.
The vast majority of these labor costs are incurred during initial implementation and have the advantage of being generally robust against other changes in LAMMPS.

As a result of the rapidly increasing complexity of the machine-learning-based potentials and their adoption of deep learning architectures, Python-based deep learning libraries such as Tensorflow, PyTorch and JAX are also becoming popular for implementing the potentials.
Potentials such as NequIP~\cite{nequip}, MACE~\cite{mace}, Allegro~\cite{allegrogb}, and HIP-NN~\cite{hipnn} use these Python-based deep learning libraries along with Kokkos.
These libraries provide auto-differentiation, which eliminates the need for manual derivative implementations, hardware portability, and pre-existing implementations of common building blocks.

LAMMPS supports two integration strategies for potentials implemented using these Python packages.
The first is to use the C++ interface to the libraries for those that have one, such as the \texttt{libtorch} interface to PyTorch.
This requires linking LAMMPS to the corresponding library, having the library be linked to versions of GPU libraries such as CUDA that are compatible with the ones that Kokkos is being linked to, etc., but otherwise these potentials do not require any modifications to LAMMPS itself. 
This strategy has been adopted by NequIP~\cite{nequip}, MACE~\cite{mace}, and others, with Allegro~\cite{allegrogb} being the first to combine PyTorch with Kokkos to avoid CPU-GPU data transfer.
The second strategy is to embed a Python interpreter in LAMMPS and use it to call the Python libraries, which has the advantage of a potentially simpler installation process.
Furthermore, it allows the use of Python packages like JAX which lack a C++ interface altogether.
The ML-IAP package in LAMMPS supports this strategy with optional Kokkos acceleration, and is used successfully by HIP-NN~\cite{hipnn}.

\section{ReaxFF Exascale Preparedness: Integer Overflow}\label{sec:reaxint}

In the approach to the exascale, and at the exascale, we have been able to run much larger global problems \textit{with} much larger local sub-domains per MPI rank. 
When running on the device there is typically a one-to-one mapping between MPI ranks and GPUs\footnote{In the case of MI250X, there is one MPI rank per GCD, and for PVC there is one MPI rank per stack.}. 
We can now more easily hit numbers that overflow 32-bit integers, for example the global number of atoms can be greater than $\approx$2 billion, and data structures on each GPU can become much larger than $\approx$2 GB in size.

In some cases these phenomena were seen coming or had already been relevant, such as with global atom counts or the size of neighbor tables. LAMMPS supports a compile-time abstraction where some quantities typed as \texttt{bigint} can be promoted to 64-bit integers when the user knows their simulations may push this limit.
In the case of ReaxFF, 32-bit integer overflow was detected and fixed as part of tests at scale by replacing hard-coded \texttt{int} with \texttt{bigint}. However, other cases required non-trivial refactors. 
We will describe two cases below.

The first case is with the sparse matrix-vector storage format. 
The number of non-zero values per MPI rank can exceed two billion for sufficiently large local systems. 
One solution to this is to promote all integer data structures---row offsets, column indices, and row lengths---to 64-bit integers. 
However, this is needlessly wasteful: the column indices and row lengths are bound by the \textit{rank} of the matrix, it is only the row offsets that will exceed the range of 32-bit integers on any reasonable system. 
For this reason the row offsets data structure, which is of length $N_\text{atoms}$, where $N_\text{atoms}$ is the number of atoms owned by an MPI rank, was promoted to 64-bit integers. 
The other two integer data structures---most significantly the column indices, which is of length $\approx N_\text{atoms} \times N_\text{neigh}$, where $N_\text{neigh}$ is the maximum number of neighbor atoms---can remain a more space-efficient 32-bit integer View.

The second case was with the bond order and hydrogen bond neighbor table. 
These were initially implemented as flat 1-d Views where 32-bit integer overflow occurred when indexing into the 1-d View. 
This was not a trivial refactor because integer offsets into the flat 1-d View were packed into \texttt{int4} data structures described in Section~\ref{sec:reaxpreprocessing}; promoting that to 64-bit integers may have been prescriptive but non-trivial. 
The more robust solution was to replace the flat 1-d Views with more natural 2-d neighbor tables. 
Here no index exceeded a 32-bit integer. 
This still required a code refactor; routines needed to switch from encoding the full 32-bit offset to the neighbor index offset. 
However this avoided the need to needlessly promote data structures to 64-bit integers. 

\section{Comparisons of Alps and Eos}\label{sec:eos-alps}
\definecolor{alps}{RGB}{35,139,69}
\definecolor{aurora}{RGB}{107,174,214}
\definecolor{frontier}{RGB}{239,59,44}
\definecolor{eos}{RGB}{253,141,60}
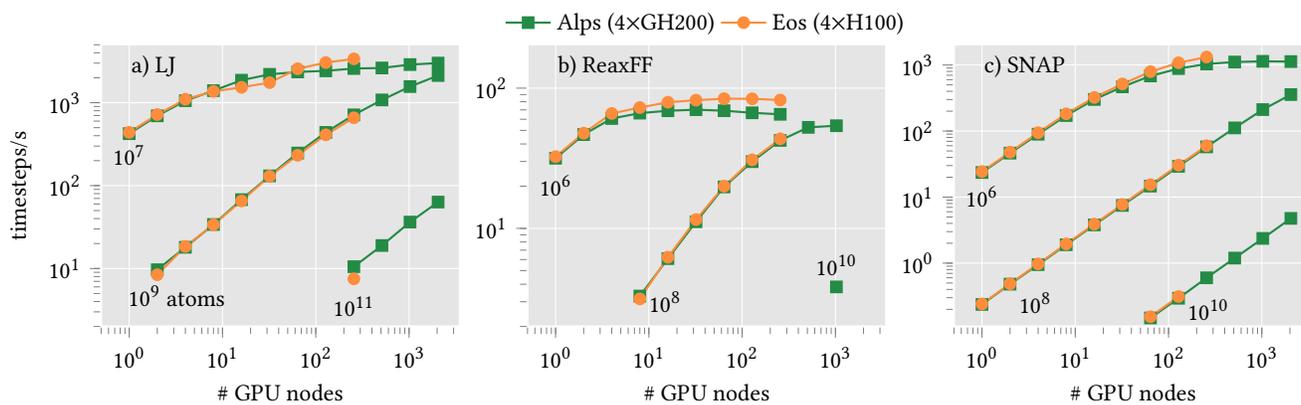
\begin{figure*}[tbp]
\if\regenfigs1
\begin{tikzpicture}[overlay=false,>=latex]
    \begin{groupplot}[
            /tikz/overlay=false,
            /tikz/thick,
            group style={
                group size=3 by 1,
                vertical sep=0.2cm,
                horizontal sep=0.9cm,
                xlabels at=edge bottom,
                ylabels at=edge left,
                xticklabels at=edge bottom,
            },
            width=\linewidth/2,
            ylabel near ticks,
            xlabel={\# GPU nodes},
            xmode=log,
            ymode=log,
            log basis x=10,
            xmin=0.5, xmax=3500,
            clip=false,
            title style={yshift=-0.5cm, xshift=-2.1cm,anchor=west},
            ylabel={timesteps/s},
            height=2.1in,
            width=2.5in,
        ]
            
            \nextgroupplot[
                ymin=2, ymax=5050,
                legend style={at={(0.7,0.3)},anchor=east},
                legend columns=1,
                title={a) LJ},
            ]
            
            \addplot+[mark=square*,alps]  table[y expr={\thisrowno{2}*0.8192}] {results/strongscaling/results_alps/lj-8M-4gpus.txt};
            \addplot+[mark=square*,alps]  table[y expr={\thisrowno{2}*1.05}] {results/strongscaling/results_alps/lj-1B-4gpus.txt};
            \addplot+[mark=square*,alps]  table[y expr={\thisrowno{2}*1.342}] {results/strongscaling/results_alps/lj-128B-4gpus.txt};

            \addplot+[mark=*,eos]  table[y expr={\thisrowno{2}*0.8192}] {results/strongscaling/results_eos/lj_8M.txt};
            \addplot+[mark=*,eos]  table[y expr={\thisrowno{2}*1.05}] {results/strongscaling/results_eos/lj_1B.txt};
            \addplot+[mark=*,eos]  table[y expr={\thisrowno{2}*1.342}] {results/strongscaling/results_eos/lj_128B.txt};
            
            \draw (axis cs:1,400) node[anchor=north] {\(10^7\)};
            \draw (axis cs:2,8) node[anchor=135] {\(10^9\) atoms};
            \draw (axis cs:256,6) node[anchor=north] {\(10^{11}\)};
            
            \nextgroupplot[
                ymin=2, ymax=205,
                legend style={at={(0.5,1.0)},anchor=south},
                legend columns=9,
                title={b) ReaxFF},
            ]
            \addplot+[mark=square*,alps]  table[y expr={\thisrowno{2}*0.9339}] {results/strongscaling/results_alps/reaxff-1M-4gpus.txt};
            \addlegendentry{Alps (4\(\times\)GH200)};
            \addplot+[mark=*,eos]  table[y expr={\thisrowno{2}*0.9339}] {results/strongscaling/results_eos/hns_1M.txt};
            \addlegendentry{Eos (4\(\times\)H100)};
            
            \addplot+[mark=square*,alps]  table[y expr={\thisrowno{2}*1.195}] {results/strongscaling/results_alps/reaxff-128M-4gpus.txt};
            \addplot+[mark=square*,alps]  table[y expr={\thisrowno{2}*1.53}] {results/strongscaling/results_alps/reaxff-16B-4gpus.txt};
            \addplot+[mark=*,eos]  table[y expr={\thisrowno{2}*1.195}] {results/strongscaling/results_eos/hns_128M.txt};
            
            \draw (axis cs:1,28) node[anchor=north] {\(10^6\)};
            \draw (axis cs:8,3) node[anchor=west] {\(10^8\)};
            \draw (axis cs:1024,4) node[anchor=south] {\(10^{10}\)};
            
            \nextgroupplot[
                ymin=0.11, ymax=2050,
                legend style={at={(1.0,0.2)},anchor=east},
                legend columns=1,
                title={c) SNAP},
            ]
            
            \addplot+[mark=square*,alps]  table[y expr={\thisrowno{2}*1.024}] {results/strongscaling/results_alps/snap-1M-4gpus.txt};
            \addplot+[mark=square*,alps]  table[y expr={\thisrowno{2}*1.31}] {results/strongscaling/results_alps/snap-128M-4gpus.txt};
            \addplot+[mark=square*,alps]  table[y expr={\thisrowno{2}*1.68}] {results/strongscaling/results_alps/snap-16B-4gpus.txt};
            
            \addplot+[mark=*,eos]  table[y expr={\thisrowno{2}*1.024}] {results/strongscaling/results_eos/snap_1M.txt};
            \addplot+[mark=*,eos]  table[y expr={\thisrowno{2}*1.31}] {results/strongscaling/results_eos/snap_128M.txt};
            \addplot+[mark=*,eos]  table[y expr={\thisrowno{2}*1.68}] {results/strongscaling/results_eos/snap_16B.txt};
            
            \draw (axis cs:1,20) node[anchor=north] {\(10^6\)};
            \draw (axis cs:2,0.25) node[anchor=west] {\(10^8\)};
            \draw (axis cs:128,0.2) node[anchor=west] {\(10^{10}\)};
    \end{groupplot}
\end{tikzpicture}
\else
\includegraphics{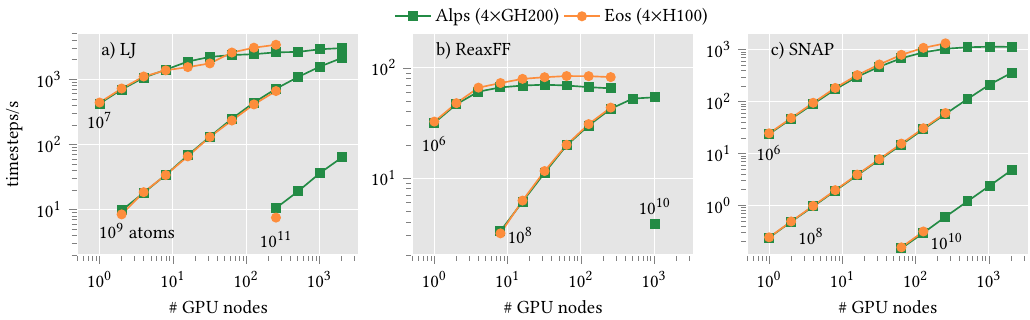}
\fi
\caption{Strong scaling results for LAMMPS on NVIDIA Eos and CSCS Alps, comparing H100 to GH200 performance across all benchmarks. The data from Alps is identical to the data from figure~\ref{fig:strongscaling}.}\label{fig:eosalps}
\end{figure*}

In figure~\ref{fig:eosalps} we show performance comparisons of Lennard-Jones, ReaxFF, and SNAP between the Alps supercomputer at CSCS~\cite{alps} and NVIDIA's Eos DGX Superpod~\cite{eos}. We deferred discussion of these machines because of their strong similarity relative to Frontier, El Capitan, and Aurora.

Each node of Alps features four Grace-Hopper superchips (GH200) and Slingshot-11 network in a 1:1 GPU to NIC ratio. Each node of Eos is a DGX H100 featuring eight GPUs per node and NDR 400 for connectivity, again with a 1:1 GPU to NIC ratio. We \textit{intentionally} only used four GPUs and four NICs per node to match the node configuration of Alps.

Exact hardware differences between H100 and GH200 are enumerated in table~\ref{tab:gpu-specs}. The relevant differences between the GPUs is that GH200 has a 20\% higher memory bandwidth and capacity than H100; while not noted on the table, it also has 20\% higher L2 capacity (50 MiB vs 60 MiB) and a commensurately higher peak L2 throughput. The key similarity is the FP64 performance and unified cache capacities are the same. These similarities and differences lead to different effects on each benchmark, including different behaviors depending on the degree of strong scaling.\footnote{Since these benchmarks are fully GPU offloaded, the high CPU-GPU bandwidth on Grace-Hopper relative to PCIe on x86+Hopper has a negligible effect on performance, as does the differences between the Grace CPU on Alps and the Xeon Platinum 8480C 56C on Eos (latency effects aside, as described in the text).}

\subsection{Lennard-Jones}

We see that when we are not deep into the strong-scaling regime, i.e. larger atom-per-GPU problems, Lennard-Jones achieves higher performance on GH200 than H100. This is because the force calculation is L2 throughput limited as opposed to compute bound. This is due to the low compute intensity of the Lennard-Jones potential.

On the other hand, in the deep strong-scaling regime, Eos with H100 outperforms Alps with GH200. This is because of higher launch latencies on GH200 which is more exposed at small per-GPU problem sizes. One approach to amortize these costs is to ``reverse'' offload some work back to the CPU, inherently removing launch latencies.

This is exposed in LAMMPS via the ``\texttt{-pk kokkos pair/only on}'' command-line argument, or by manually modifying the suffixes of certain styles so they run on the CPU instead of the GPU (see section~\ref{sec:exe-control}). We did not explore this tuning space as part of these investigations to simplify the comparison, both between H100 and GH200 and separately with MI250X, MI300A, and PVC.

\subsection{ReaxFF}

For large per-GPU problem sizes, we see a broad similarity in performance between H100 and GH200. This is owing to many of the dominant kernels being compute-limited. One potential exception is the sparse matrix-vector kernel, which in an ideal implementation should be memory bandwidth limited. Profiling indicates that this kernel is limited by latency effects due to the pointer indirection inherent to CSR-like matrix formats. Optimization to remove this limiter is an avenue of further work orthogonal to this paper.

In the deep strong-scaling regime we see Eos outperforming Alps. This is again because of latency effects. ReaxFF has more susceptibility to latency effects because of the several packing and unpacking kernels in the distributed sparse matrix-vector kernel.

\subsection{SNAP}

The top kernels of the SNAP potential are all either FP64 limited or L1 throughput limited. The performance of each is identical between H100 and GH200. In addition, communications are negligible as a percentage of end-to-end runtime, owing to the high degree of computation in the SNAP force evaluation. For these reasons, performance on Alps and Eos is very close.

\section{Reproducibility}
Details needed to reproduce the results in this work, including compiler versions, can be found at the following GitHub repository: https://github.com/megmcca/p3hpc\_ad\_lmp-kk 

\end{document}